%% file: main.tex
\documentclass[10pt]{article}

\input{header}

\algrenewcommand\algorithmicrequire{\textbf{Inputs:}}
\algrenewcommand\algorithmicensure{\textbf{Outputs:}}

\allowdisplaybreaks

\title{Measuring capacities in multimodal maritime port systems with anchorage queues}

\author[a]{Debojjal Bagchi\thanks{Corresponding author: Email: debojjalb@utexas.edu, Mobile: +1 (737) 304-9684.\\
Email addresses: debojjalb@utexas.edu (Debojjal Bagchi), bathgate@utexas.edu (Kyle Bathgate), sboyles@austin.utexas.edu (Stephen D. Boyles), kenneth.n.mitchell@usace.army.mil (Kenneth N. Mitchell), magdalena.asborno@usace.army.mil (Magdalena I. Asborno), marin.m.kress@usace.army.mil (Marin M. Kress)}}
\author[a]{Kyle Bathgate}
\author[b]{Kenneth N. Mitchell}
\author[c]{Magdalena I. Asborno}
\author[b]{Marin M. Kress}
\author[a]{Stephen D. Boyles}

\affil[a]{\small Fariborz Maseeh Department of Civil, Architectural and Environmental Engineering, The University of Texas at Austin, 301 E. Dean Keeton St., Austin, TX, 78712, USA}
\affil[b]{\small Coastal and Hydraulics Laboratory, US Army Corps of Engineers Research and Development Center, 3909 Halls Ferry Rd., Vicksburg, MS, 39180, USA}
\affil[c]{\small Applied Research Associates, Inc. for US Army Corps of Engineers Research and Development Center, 3909 Halls Ferry Rd., Vicksburg, MS, 39180, USA}

\date{}

\begin{document}
\maketitle
\vspace{-5mm}
\begin{abstract}

This paper presents a framework for estimating the capacity of a multimodal maritime port system handling vessels of multiple classes. Port system capacity can be categorized into two distinct types: operating capacity, defined as the maximum number of vessels that can be processed over an extended period under stable operating conditions, and ultimate capacity, defined as the absolute maximum vessel throughput achievable irrespective of stability. Distinguishing between these two capacity measures is critical for long-term planning and resilience analysis, as ports may temporarily operate above sustainable levels following disruptions or during demand surges. Despite the importance of this distinction, existing port capacity models generally do not provide methods to compute port-level capacity estimates that clearly differentiate between operating and ultimate capacity. We introduce methods to estimate both capacity measures for seaport systems. We apply the proposed framework using the Port of Houston, Texas as a case study. Operating capacity is estimated using a parsimonious queueing-theoretic model, while ultimate capacity is estimated by fitting an ordinary differential equation model to simulation outputs. We estimate an operating capacity of approximately $0.9$~vph and an ultimate capacity of approximately $1.4$~vph for the Port of Houston. Sensitivity analysis of key port resources indicates that liquid-bulk terminals constitute the primary bottlenecks under stable operating conditions, whereas pilot availability becomes the dominant bottleneck following disruptions. These methods can be used in port planning to determine the expected operational and resilience gains of a given infrastructure intervention, or to identify bottlenecks in a complex, multimodal port environment.

\vspace{3mm}
\end{abstract}

\section{Introduction} \label{sec:intro}

Measuring the capacity of multimodal port systems is essential for identifying bottlenecks, guiding infrastructure investment, and supporting recovery planning following disruptions. Seaports handle over 80\% of global trade by volume \citep{unctad2023review}; however, unlike roadway systems, ports lack standardized methods for estimating system-level capacity. System-level capacity measures are particularly valuable when ports are modeled as nodes within larger freight transportation networks and are critical for informing port-level decision making by public-sector agencies. 

When port infrastructure cannot immediately accommodate arriving vessels, ships wait at designated offshore areas known as anchorages. The resulting accumulation of vessels forms an anchorage queue, which captures system-level congestion arising from interactions among waterway channels, terminals, pilots, and landside infrastructure. As such, anchorage queues provide a natural and observable indicator of port system capacity. We argue that port system capacity must be understood through two distinct concepts. The first is the maximum vessel throughput that can be sustained over an extended period under stable operating conditions. The second is the maximum throughput that can be achieved over a finite time horizon, irrespective of operational stability. We refer to the former as \textit{operating capacity} and the latter as \textit{ultimate capacity}. 

The applications of operating capacity and ultimate capacity are distinct. Operating capacity should be used to study long-term or steady-state port behavior, as it reflects the maximum sustainable vessel throughput under stable conditions. In contrast, ultimate capacity is appropriate when information about the port’s maximum processing rate is needed irrespective of stability, such as post-disruption recovery operations or during temporary demand surges. As we show later in our results, different resources can become bottlenecks under different circumstances: terminal berths may restrict flow during normal operations, whereas the number of available pilots may become the binding constraint during recovery. 

\begin{figure}
  \centering
  \begin{subfigure}[]{0.45\textwidth}
    \centering
    \includegraphics[width=\textwidth]{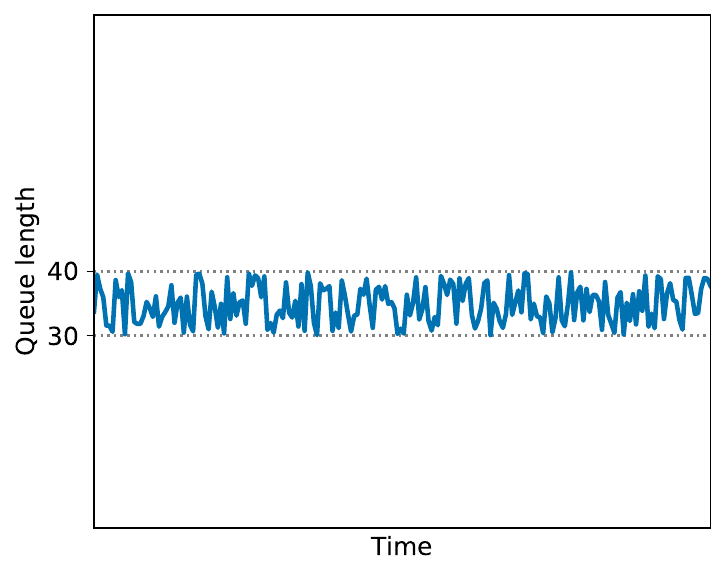}
    \caption{Stable queue.}
    \label{fig:stable}
  \end{subfigure}
  \hfill
  \begin{subfigure}[]{0.45\textwidth}
    \centering
    \includegraphics[width=\textwidth]{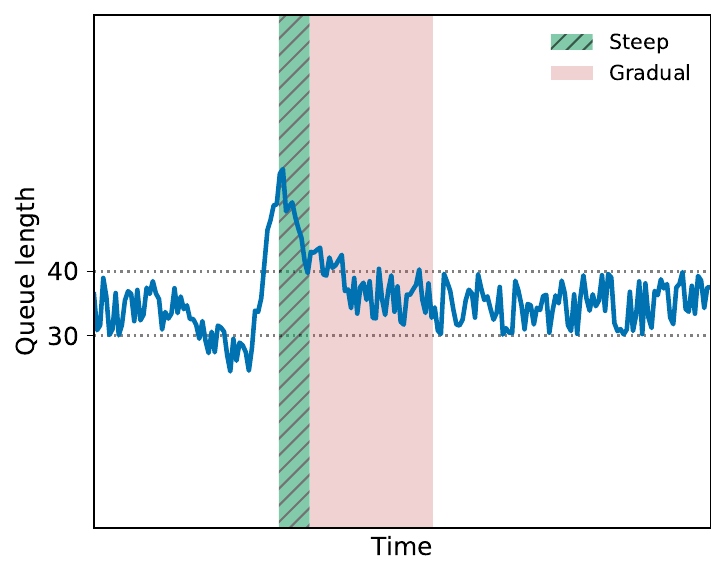}
    \caption{Unstable queue.}
    \label{fig:unstable}
  \end{subfigure}
  \caption{Different anchorage queue dynamics over time.}
  \label{fig:queue_dynamics}
\end{figure}

To illustrate the distinction between operating and ultimate capacity, consider Figure~\ref{fig:queue_dynamics}, which shows typical anchorage queue behavior under both stable and unstable conditions. Figure~\ref{fig:stable} depicts stable operations, where queue lengths fluctuate steadily between 30 and 40 vessels over an extended period. In contrast, Figure~\ref{fig:unstable} shows a port closure followed by recovery. Such closures may occur during tropical storms, periods of heavy fog, or similar operational disruptions. During a closure, the anchorage queue grows significantly as all new vessel arrivals join the queue while no vessels are processed out of the anchorage. Once operations resume, the port can temporarily accommodate these elevated queues through brief periods of higher processing rates. However, these high-capacity conditions are short-lived and represent an \textit{unstable system}. The port system eventually returns to its usual operating capacity as conditions stabilize. This behavior is highlighted by the two shaded regions in Figure~\ref{fig:unstable}. The steep initial decline corresponds to a phase of processing near ultimate capacity, whereas the more gradual decline reflects the system’s return to operating capacity.

\subsection{Research gap}

To the best of the authors' knowledge, no prior studies in port modeling have proposed a classification between operating and ultimate capacities or developed methods to compute and compare both. Previous research has attempted to define port capacity either based on individual components \citep{salminen2013measuring} or through simulation approaches \citep{bellsola2017network}, but both approaches face notable limitations.

\citet{bellsola2017network} argue that the capacity of a port network cannot be determined by independently estimating the capacities of individual components and then selecting the smallest one, because these components are interdependent. A resource with a low nominal service rate may not constitute a true bottleneck if it receives only a small share of the overall traffic. For instance, consider a waterway channel that can handle one vessel per hour feeding a terminal whose crane can serve one vessel every two hour. If only one-tenth of the channel’s flow is destined for that terminal, the crane does not constrain system throughput because it never receives enough arrivals to reach its service limit. This example shows that, while the nominal capacities of cranes, berths, or pumps may be simple to estimate, determining the overall port capacity requires understanding the port system as a whole. 

Conversely, current simulation-based approaches lack a theoretical foundation for defining capacity. Instead, they rely on proxy indicators, such as the ratio of waiting time to service time, vessel turnaround time, berth occupancy rate, or the total number of vessel trips completed within a given period. Furthermore, in practice, ports may occasionally handle higher vessel volumes within short time intervals, but such conditions are unstable and cannot be maintained over extended periods. Most studies overlook the stability of these indicators, even though it plays a critical role in determining whether the estimated capacities represent sustainable or temporary conditions. 

Another limitation of current port capacity assessment methods is their dependence on extensive, high-quality data, which is frequently difficult to obtain due to operational complexity and restricted access. Even when data are available, they often lack the granularity or fidelity required to accurately calibrate simulation or interaction models. Consequently, there remains a significant research gap in developing models capable of estimating port capacity under varying levels of data availability. 

\subsection{Contributions}

In this paper, we present methods to compute the operating and ultimate capacities of seaport systems. Our models are applicable to any port with an open anchorage followed by a waterway channel leading to terminals. Even when port structures differ from the one studied here, the methods can be easily adapted to specific port configurations. Operating capacity is computed using a queueing theory-based approach, while ultimate capacity is computed using an ordinary differential equation (ODE)-based model. As a result, operating capacity can be estimated using only archival Automatic Identification System (AIS) vessel tracking data and port records, whereas ultimate capacity requires a simulation of the port, thereby allowing capacity estimation under different levels of data availability. In summary, our work makes four main contributions to the literature. The first three are methodological advancements, while the fourth demonstrates the practical applicability of the proposed framework:
\begin{enumerate}
\item \textbf{Queueing theory-based model for operating capacity:}
We present a parsimonious queueing theory-based model for estimating the operating capacities of port systems under limited data availability. Queueing theory models have been applied in other domains, such as healthcare \citep{lantz2016measuring, bittencourt2018hospital} and logistics \citep{annas2022developing} to measure system capacities. However, these models do not directly apply to port operations. We develop queueing models tailored to port operations to compute the operating capacity of a seaport system. Our proposed model captures long-term port behavior while incorporating level-of-service parameters such as queue lengths and waiting times. The proposed model can estimate port capacities directly from archival AIS vessel-tracking data and terminal records, without relying on simulations. Since the model does not require simulation, it serves as an ideal initial screening tool for determining whether a detailed capacity assessment is necessary. Unlike existing port capacity estimates, our operating capacity model focuses on long-term sustainable throughput without relying on peak throughput observed in simulations, which is often unstable. This makes the model suitable for long-term infrastructure planning and bottleneck identification under stable operating conditions.

\item \textbf{Simulation-driven ODE model for ultimate capacity:}
To estimate the ultimate capacity of a port, we introduce a ODE-based model to predict the number of vessel processed by the port as a function of arrival rate. Although not used for estimating port capacity, ODE-based models have long been applied to represent the carrying capacity of biological populations \citep{ meyer1999carrying, del2004carrying}. Our framework uses a detailed simulation of the port to fit a dynamic model that estimates the port's ultimate capacity. Unlike prior simulation-based port capacity studies \citep{bellsola2017network, chen2013simulation}, which rely on statistical curve fitting or performance indicators, our model provides a theoretical foundation for defining capacity based on simulated system responses. The resulting ultimate capacity estimates aid stakeholders in post-disruption planning by identifying bottlenecks and determining the maximum recovery rate following disruptions and port-closures.

\item \textbf{Link between operating and ultimate capacity:}
We formally establish a relationship between operating and ultimate capacity. Although these metrics arise from different formulations, we show that operating capacity is always strictly less than ultimate capacity and that operating capacity can be inferred from the ultimate capacity model. We further identify conditions under which the two can become arbitrarily close. We clarify the distinction between these capacity measures and provide guidance on which metric is most appropriate for different decision-making contexts. The distinction between operating and ultimate capacities indicate maximizing throughput in a port can lead to future congestion.  

\item \textbf{Case study of the Port of Houston:} We validate our framework through a data-driven discrete event simulation calibrated using archival AIS and operational data from the Port of Houston. The case study demonstrates the framework’s ability to detect bottlenecks through sensitivity analyses of key operational resources and illustrates how the proposed metrics can support decision-making across diverse port settings. The complete simulation framework, including capacity estimation procedures, is released as open-source software for use by port practitioners and researchers\footnote{The simulation developed in this study is available at \url{https://spartalab.github.io/port-simulation/}. Due to data confidentiality restrictions, the input data corresponding to the Port of Houston are not provided. However, representative sample input data for a model port are included to enable testing and replication of the methods proposed in this study.}.
\end{enumerate}

\subsection{Organization}

The remainder of the paper is structured as follows. Section~\ref{sec:litrev} reviews the relevant literature on port capacity. Section~\ref{sec:meth} motivates the need for two capacity notions and clarifies their practical distinctions. Section~\ref{subsec:op_cap} introduces the queueing theory-based operating capacity model, while Section~\ref{subsec:ul_cap} presents the ODE-based ultimate capacity model and its relationship with operating capacity. Section~\ref{sec:results} outlines our simulation procedure and evaluates the proposed methods using empirical and simulated data from the Port of Houston. Section~\ref{sec:appnl} discusses practical applications of the models for port planning and decision-making. Finally, Section~\ref{sec:conc} summarizes the key findings and identifies directions for future research.

\section{Literature} \label{sec:litrev}

The Bureau of Transportation Statistics (BTS) \citep{BTS2024} defines port capacity as a “measure of the maximum throughput in tons, twenty-foot equivalent units (TEU), or other units that a port and its terminals can handle over a given period.” However, measuring capacity in a multi-modal port system is inherently challenging due to the complex interactions among multiple resources and parameters \citep{vasheghani2023strategic}. Consequently, there is no universally accepted method for computing port capacity \citep{bellsola2017network}. As discussed in Section \ref{sec:intro}, capacity computations in seaports typically stem from two main domains, component-based and simulation-based approaches.

When the capacity of a seaport system is computed based on the capacities of individual components, two definitions are commonly found in the literature, static and dynamic capacity. Static capacity refers to the volume a port can handle at a specific point in time and is determined by space availability, while dynamic capacity refers to the volume a port can handle over a period of time \citep{lagoudis2011revisiting}. \citet{lagoudis2011revisiting} and \citet{salminen2013measuring} differentiated between static and dynamic capacities when defining capacities for several port components. Static capacity in ports corresponds to what is often termed “jam density” in roadway networks, representing the number of vessels that can be accommodated within a fixed space. For example, \citet{liu2022research} estimated the number of vessels that can be safely accommodated in navigable waters to compute waterway carrying capacity. In contrast, dynamic capacity aligns with the BTS definition, referring to the throughput that can pass through a resource over a specific period.

Numerous studies have examined the capacities of individual port components. For example, \citet{fan2000sea} defined the capacities of several seaport components, including waterway links, intersections, and berths, based on the average separation between vessels, considering capacity as the number of vessels that can be served within a given time interval. Similarly, \citet{liu2020ais} characterized waterway capacity as “the ratio of the spatiotemporal resources of the port waterway to the spatiotemporal resources occupied by a single ship generally sailing in and out of the port waterway.” At the berth level, \citet{noritake1983optimum} employed berth capacity curves to determine the optimal cargo capacity for general cargo berths, while \citet{park2014port} proposed a formula for estimating berth capacity. In a broader review, \citet{chang2012estimation} summarized several methods for estimating berth capacity in container ports.

While the capacity of individual components is important, port-level decision-making often requires an aggregate measure of capacity for the entire port. This is typically achieved by first estimating the capacities of individual components and then either combining them through interaction functions to derive overall port capacity \citep{tafur2024flow} or by identifying the most restrictive component \citep{fan2000sea}. However, in real-world operations, vessels are rarely tightly packed, and capacity is influenced by factors beyond physical dimensions, such as the availability of pilots and tugs, downstream terminal resources, and navigation restrictions. Furthermore, calculating network capacity based solely on individual components may fail to capture the interdependencies among port subsystems \citep{bellsola2017network}.

Computing the exact capacity of an entire port system, or even its individual components, is challenging due to the complex interactions among subsystems such as channels, pilotage, terminal operations, and various cargo modes. Consequently, many capacity models rely on simulations to explore different operational scenarios, identify bottlenecks, and gain insights into port performance \citep{chen2013simulation}. For example, \citet{huang2011assessing} evaluated anchorage capacity by simulating a realistic vessel mix and analyzing anchorage utilization, while \citet{o2005dynamic} determined the minimum channel dimensions required for safe navigation under various scenarios using query-and-simulate loops. Similarly, \citet{chen2013simulation} computed capacity for inland waterway networks, defining capacity based on the ratio of average waiting time to the normal passage time through the network. Although these simulation-based approaches are effective, they primarily focus on the capacities of individual components rather than the port system as a whole.

\begin{table}
\centering
\small
\renewcommand{\arraystretch}{1.25}
\setlength{\tabcolsep}{4pt}
\begin{tabular}{
|>{\raggedright\arraybackslash}p{2.0cm}|
>{\raggedright\arraybackslash}p{3.0cm}|
>{\raggedright\arraybackslash}p{1.4cm}|
>{\raggedright\arraybackslash}p{1.5cm}|
>{\raggedright\arraybackslash}p{1.9cm}|
>{\raggedright\arraybackslash}p{\dimexpr\textwidth-11cm-5\tabcolsep}|}
\hline
\textbf{Study} & \textbf{Scope} & \textbf{Data} & \textbf{Capacity type} & \textbf{Model type} & \textbf{Method} \\
\hline
\citet{fan2000sea} & Individual resources \& entire port (component-wise) & Real & Dynamic & Analytical & Defines component capacities individually and overall port capacity as the most restrictive component. \\ \hline
\citet{o2005dynamic} & Waterway & Real & Dynamic & Simulation & Query-and simulate loops to compute of capacity. \\ \hline
\citet{huang2011assessing} & Anchorage & Real & Static \& dynamic & Simulation & Space-utilization-based assessment. \\ \hline
\citet{lagoudis2011revisiting} & Individual resources & Real & Static \& dynamic & Analytical & Static and dynamic capacity definitions of individual components. \\ \hline
\citet{chen2013simulation} & Inland waterway & Real & Dynamic & Simulation & Capacity derived from ship waiting-to-passage time ratio. \\ \hline
\citet{salminen2013measuring} & Individual resources \& entire port (component-wise) & Real & Static \& Dynamic & Analytical &  Defines component capacities individually and identify the most restrictive component.\\ \hline
\citet{lee2014analysis} & Container terminal & Synthetic & Dynamic & Queueing & Terminal queueing-based capacity model, \\ \hline
\citet{bellsola2015estimating} & Entire port (integrated) & Synthetic & Dynamic & Simulation & Curve fitting to simulated waiting time to service time ratios; capacity estimated as inflection point in the curve (visual inspection). \\ \hline
\citet{souf2016port} & Container terminal & Real & Dynamic & Queueing \& simulation & Throughput, utilization, and anchorage use as qualitative capacity indicators. \\ \hline
\citet{bellsola2017network} & Entire port (integrated) & Synthetic & Dynamic & Simulation & Exponential curve fitting on the relationship between total trips and the waiting time to service time ratio. \\ \hline
\citet{liu2020ais} & Waterway & Real & Dynamic & Analytical & Ratio of available to occupied spatiotemporal waterway resources. \\ \hline
\citet{tafur2024flow} & Entire port (component-wise)& Real \& synthetic & Static \& dynamic & Analytical & Interaction functions combining component capacities; flow limited by bottleneck component. \\ \hline
\textbf{This paper} & Entire port (integrated) & Real & Dynamic & Analytical, queueing \& simulation & Integrated model to estimate maximum vessel flow through port; Distinction between operating and ultimate capacities. \\ 
\hline
\end{tabular}
\caption{Summary of key port capacity models in the literature. \label{tab:lit}}.
\end{table}

More recent studies have aimed to estimate the overall capacity of entire port systems using simulation. \citet{bellsola2015estimating} and \citet{bellsola2017network} introduced the concept of port network traffic capacity (PNTC) to represent the capacity of the overall port network. \citet{bellsola2017network} defined PNTC as “the maximum average vessel flow that can be handled by a port, with its specific infrastructure layout, vessel fleet, traffic composition, and demand, satisfying the required safety and service level.” To estimate this value, they conducted multiple simulations with varying demand levels to identify ``saturated" conditions. While their approaches align with the BTS definition of port capacity, their approaches depend purely on statistical methods to quantify ``saturation" and are not based on the physical dynamics of port operations. For example, \citet{bellsola2017network} approximated capacity by fitting an exponential curve to the total trips and the ratio of waiting-time to service-time derived from simulated vessel flow data. However, this curve-fitting approach has limited theoretical grounding, as indicated by relatively low $R^2$ values $(0.38$--$0.53)$ even in a synthetic three-terminal network. Furthermore, they do not consider the impact of the simulation time horizon on their results, leaving it unclear whether the capacity estimates reflect long-term conditions or transient, unstable states. Some of the key literature on port capacity estimation is summarized in Table \ref{tab:lit}.

Beyond capacity estimation, understanding port operations has been a significant research focus. Several studies \citep{canonaco2008queuing, dragovic2012mathematical, legato2014queuing, easa1987approximate} have applied queueing theory to model port processes. \citet{zrnic1999anchorage} modeled the anchorage-ship-berth link as a multi-server queueing system, while \citet{mrnjavac2000modelling} simulated container terminal operations using known arrival and service rates, analyzing the effect of increased arrival rates under a fixed number of berths. Queueing theory and simulation have also been used to determine the optimal number of berths \citep{el2010application} and to assess the impact of operational disruptions \citep{guo2023network}. \citet{souf2016port} combined simulation modeling with AIS data to estimate container terminal throughput, while \citet{bugaric2007increasing} analyzed strategies to enhance terminal capacity without significant investment. Landside infrastructure and port-hinterland connections have also been modeled using queueing theory. For example, \citet{chen2013reducing} estimated truck waiting times and emissions at port container terminals using queueing models for truck gates and terminal yards. At a broader level, networks of ports have been analyzed using closed Jackson network queueing models \citep{roy2018stochastic}.

Overall, the existing literature on port capacity estimation focuses on several domains. The capacity of individual port components has been computed in both static and dynamic senses, while capacity estimation at the port-system level remains significantly understudied. Port-level capacity is often inferred from individual components, which may not adequately account for interactions within a port system. Among the limited number of studies that estimate port-level capacity using simulation, such as \citet{bellsola2015estimating}, capacity is typically inferred from operational performance indicators without distinguishing whether the resulting estimates are sustainable over extended periods of operation. As a result, existing methods provide limited guidance on whether estimated capacity values represent long-run operating conditions or unstable throughput levels following short-term demand surges. Methods to compute capacities that distinguish between these two types of operational environments remains a research gap in the study of port systems. The following section addresses these limitations by introducing methods to compute these port capacities.

\section{The capacity model}\label{sec:meth}

To motivate our port capacity model, we first describe the life cycle of a vessel within a port with an open anchorage and navigation channel providing access to transload terminals. A vessel’s port life cycle begins upon entering the anchorage area, where it queues until a pilot is assigned and a berth becomes available at its destination terminal. In ports with narrow waterway channels, navigation may be constrained by beam, draft, and daylight restrictions in the channel. Once a terminal berth is available and all channel restrictions are satisfied, the vessel requests the necessary resources (such as tugboats) to proceed to its terminal berth. A vessel exits the anchorage and enters the channel only when both an available berth and the required resources are secured \citep{kang_study_2022}. Because vessels can depart the anchorage only when all berth, navigation, and resource constraints are satisfied, the rate of anchorage exits effectively represents the port’s ability to process arriving vessels. In our formulations, we therefore analyze anchorage queue dynamics and vessel entries into the channel to define port capacities. We define the following three notions of capacities in port systems. While the operating and ultimate capacities are the primary metrics for practical port planning, we introduce a third measure, saturated capacity, to completely describe the capacity model. The saturated capacity captures the behavior of the port under an extremely high arrival rate, which would not be observed in practice; however, including it in the model helps to estimate the other two capacity metrics more accurately.

\begin{dfn}[Operating capacity]
The operating capacity of a port, denoted by $C_o$, is the maximum long-term average vessel processing rate that can be sustained under stable anchorage queueing conditions.
\end{dfn}

\begin{dfn}[Ultimate capacity]
The ultimate capacity of a port, denoted by $C_u$, is the maximum rate at which vessels can be processed by a port, regardless of anchorage queue stability. 
\end{dfn}

\begin{dfn}[Saturated capacity]
The saturated capacity of a port, denoted by $C_s$, is the limiting rate at which vessels can be processed by a port as the port becomes extremely congested. 
\end{dfn}


In these definitions, port capacity is expressed in vessels per hour (vph). Using vph as the unit of throughput, rather than cargo-based measures such as volumetric flow, offers several advantages for port-level capacity analysis. First, vessel arrival rates can be directly and reliably estimated from archival AIS data, whereas detailed cargo information is often unavailable or inconsistently reported. Second, aggregating cargo flows across terminals and vessel classes requires adopting a common unit, which is difficult given the heterogeneity of cargo types handled within a port. Further, a vph-based capacity enables direct estimation of performance measures such as queue growth, recovery time following surges in vessel arrivals, and delays experienced at port under varying demand levels. Nevertheless, the methods presented in this study are applicable to any cargo flow unit, provided that all quantities, such as arrival rates and queue lengths, are expressed using consistent units.

\begin{figure}
    \centering
    \includegraphics[width=0.9\linewidth]{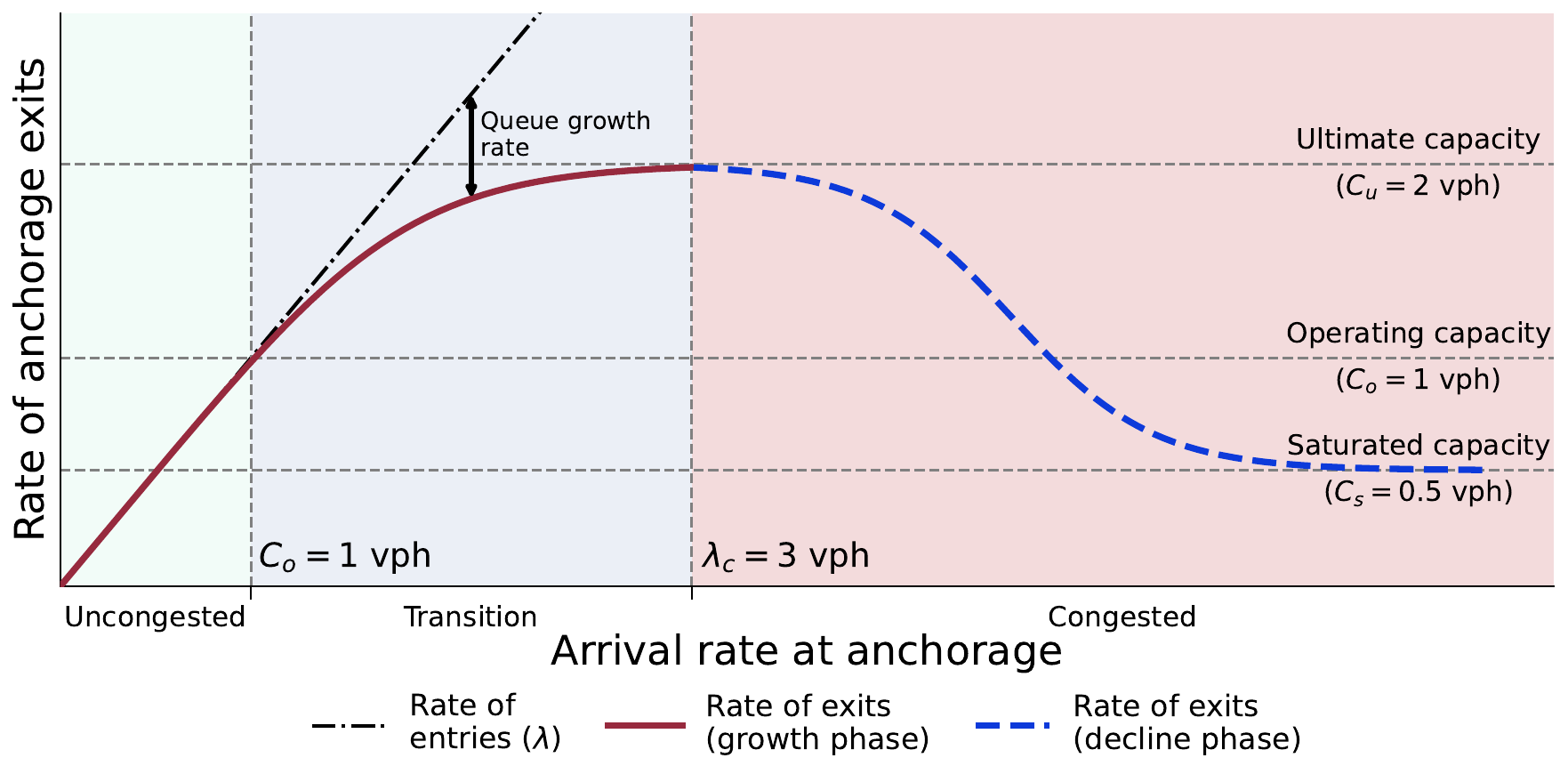}
\caption{Capacity regimes (not to scale). The figure illustrates how the rate of anchorage exits changes with the arrival rate. In the uncongested phase, the rates of entry and exit are equal, reflecting sustainable conditions. Beyond the uncongested phase, the rate of exits falls below the rate of entries, causing persistent queue growth over time. The operating, ultimate, and saturated capacities are marked.}
    \label{fig:cap_regimes}
\end{figure}

We now explain the expected behavior of anchorage exits under varying vessel arrival rates at the port. We define an \textit{exit curve} as the relationship between the number of vessel exits from the anchorage and the vessel arrival rate at the port. Figure~\ref{fig:cap_regimes} illustrates the expected typical shape of the exit curve for a port system. 

When arrival rates are low, anchorage queues may form due to stochastic arrivals, but over a given time horizon the number of vessels arriving equals the number processed. For example, if a port’s operating capacity is $1$~vph, then at an arrival rate of $\lambda = 0.9$~vph, vessels also depart at $0.9$~vph. For any arrival rate below the operating capacity, throughput increases linearly with $\lambda$, since in this regime the number of anchorage exits is simply $N^{\text{exits}} = \lambda \tau$, where $\tau$ is the observation horizon. We refer to the range of arrival rates from zero up to the operating capacity as the uncongested regime. As shown in Figure~\ref{fig:cap_regimes}, the arrival and exit rates remain close throughout this regime. 

As arrivals increase beyond the operating capacity, the processing rate by the port falls below the arrival rate. This marks the onset of the \textit{transition regime}.  Early arrivals wait less at the anchorage, while later arrivals experience longer delays as queues grow with time. The queue growth rate is the difference between arrival rate and rate at which vessels exits the anchorage (see Figure~\ref{fig:cap_regimes}). Although delays increase, the port can temporarily process more vessels than its operating capacity, reflecting its ability to absorb short-term excess demand. As arrivals continue to increase, the port throughput would reach a maximum. We refer to this maximum processing rate as the ultimate capacity of the port and denote the arrival rate at which this is achieved as $\lambda_c$. For example, if vessels arrive at a rate of $\lambda = 3$~vph in the previous example, approximately 72 vessels reach the anchorage in one day. Because the arrival rate exceeds the operating capacity, all 72 vessels cannot depart within the day. If 48 vessels exit in one day (corresponding to a rate of $2$~vph), the anchorage queue is unstable, yet the processing rate remains higher than the operating capacity. We refer to the entire range from zero arrival rate up to $\lambda_c$ as the \textit{growth phase}, as shown in Figure~\ref{fig:cap_regimes}. In this phase, increases in vessel arrival rates lead to higher port throughput.

Arrival rates exceeding $\lambda_c$ define the \textit{congested regime}. Beyond $\lambda_c$, further increases in vessel arrivals no longer increase port throughput. In fact, higher arrival rates may reduce throughput due to congestion and competition for shared resources. As congestion intensifies, the exit rate declines and eventually plateaus at the saturated capacity, denoted by $C_s$. The saturated capacity represents the minimum port throughput achievable under extreme congestion. Depending on the system, $C_s$ may be zero, indicating a complete halt in operations, or positive if some throughput is maintained. Unlike road networks, which can experience gridlock and a resulting zero-flow state, port systems are less prone to a complete collapse in throughput due to extreme congestion.  Because vessel queues form only at the anchorage, the system can maintain a non-zero $C_s$ even under exceptionally high arrival rates. Although arrival rates sufficient to observe $C_s$ are rarely encountered in practice, defining this parameter is essential for accurately characterizing the \textit{exit curve}.

We observe that the typical shape of this exit curve is qualitatively similar to the macroscopic fundamental diagrams (MFD) described by \citet{daganzo2007urban}, which relates aggregate accumulation to throughput. Specifically, our growth and decline phases are analogous to the rising and declining branches of the MFD. We also draw parallels with the MFD in that both concepts contain congested and uncongested regimes and a maximum service rate followed by service that decreases with congestion.

This discussion highlights the distinction between operating and ultimate capacity. When vessel arrivals exceed the operating capacity, the anchorage queue becomes unstable, meaning that the number of vessels waiting increases over time. Although the port may temporarily process vessels at rates above its operating capacity, up to the ultimate capacity, vessels arriving later experience substantially longer waiting times than those arriving earlier. After a port closure, when many vessels attempt to enter simultaneously, the system may initially operate near this peak rate, assuming the closure was temporary and no physical damage occurred. However, as the backlog grows, the processing rate must eventually be reduced to reestablish stable queueing conditions.

\begin{table}
\centering
\begin{tabular}{p{3cm}|p{6cm}|p{6cm}}
\hline
\textbf{} & \textbf{Operating capacity ($C_o$)} & \textbf{Ultimate capacity ($C_u$)} \\
\hline
Definition & The maximum throughput of vessels that can be sustained by the port for a long period of stable anchorage queue formation &  The maximum throughput of vessels that can be sustained by the port irrespective of anchorage queue stability conditions\\
Anchorage queue & Bounded & Increases linearly with time \\
Model & Queueing model & Simulation with ODE-based model \\
Inputs & 
Archival AIS vessel tracking data and terminal logs
& 
A detailed simulation of the port\\
Interpretation & Long term sustainable port throughout
& 
Maximum possible throughput \\
Application & Infrastructure planning, bottleneck analysis in long term planning & Post disruption planning,  bottleneck analysis post port closure\\
\hline
\end{tabular}
\caption{Operating and ultimate capacity model. \label{tab:summary}}
\end{table}

In the subsequent sections, we present methods to quantify operating and ultimate capacities, leading to a characterization of the exit curve. To this end, we develop both system level queueing and detailed simulation models. The operating capacity is derived using an aggregated queueing formulation of the anchorage-channel-terminal system, where the service rate represents the long-term sustainable throughput. The ultimate capacity is estimated through a detailed discrete-event simulation of the entire port and using an ODE-based model that describes the number of vessel departures as a function of the arrival rate. Table~\ref{tab:summary} summarizes the data requirements, modeling frameworks, outputs, and applications for both capacity measures. Detailed formulations for computing the operating capacity are provided in Section~\ref{subsec:op_cap}, and methods for determining the ultimate capacity are presented in Section~\ref{subsec:ul_cap}.

\section{Operating capacity} \label{subsec:op_cap}
Queueing theory provides an ideal framework for modeling long-term stable conditions. Moreover, by analyzing long-term stationary behavior, queueing models can yield closed-form expressions for quality-of-service measures such as queue length and waiting times, reducing reliance on detailed simulation models.  We, therefore, use queueing models of the port to define its operating capacity. In this section, we present the queueing model used to model port operations, describe how the models are applied to obtain the operating capacity, and provide guidelines for interpreting the model outputs.

Queueing models rely on a well-defined abstraction of the processes and services observed in real-world port operations. In particular, they require knowledge of arrival processes and queueing disciplines. Many studies model vessel arrivals at anchorages and terminals as a Poisson process \citep{de2020new, shabayek2002simulation}. However, empirical evaluations of arrival data \citep{legato2020queueing} suggest that this assumption does not always accurately reflect real-world port behavior, particularly for certain vessel classes. For instance, \citet{bathgateAIS} finds that while the Poisson process provides a reasonable approximation when arrivals are aggregated across all vessel classes, tanker arrival counts frequently fail standard goodness-of-fit tests for a Poisson distribution. Since detailed data on vessel arrivals at port anchorages can be obtained using AIS vessel tracking data \citep{NOAA_AIS}, both the mean and standard deviation of arrival rates can be computed directly. Hence, we adopt a general distribution to model vessel arrivals. We assume vessels of different classes (indexed by $i$) arrive at the anchorage with arrival rate $\lambda_i$ and coefficient of variation of inter-arrivals is $c_i$.

Once vessels arrive at the anchorage, they queue until all downstream terminal resource and channel restrictions are satisfied. We assume vessels wait at the anchorage independently due to both terminal unavailability and channel constraints. Since the channel does not differentiate among vessel classes, we assume a common mean waiting time due to pilot restrictions, pilot availability, and other waterway resources, denoted by $W^{C}$. The waiting time due to terminal resources, however, is class-dependent and is denoted by $W_i^{T}$. We denote the queue length of each vessel class at the anchorage due to channel constraints as $L^{C}_{i}$, and that due to terminal constraints as $L^{T}_{i}$. Our model assumes that queues form only at the anchorage, which is typical in many port systems where the anchorage is followed by a narrow channel leading to the terminals.

\begin{figure}[!h]
    \centering
    \includegraphics[width=\textwidth]{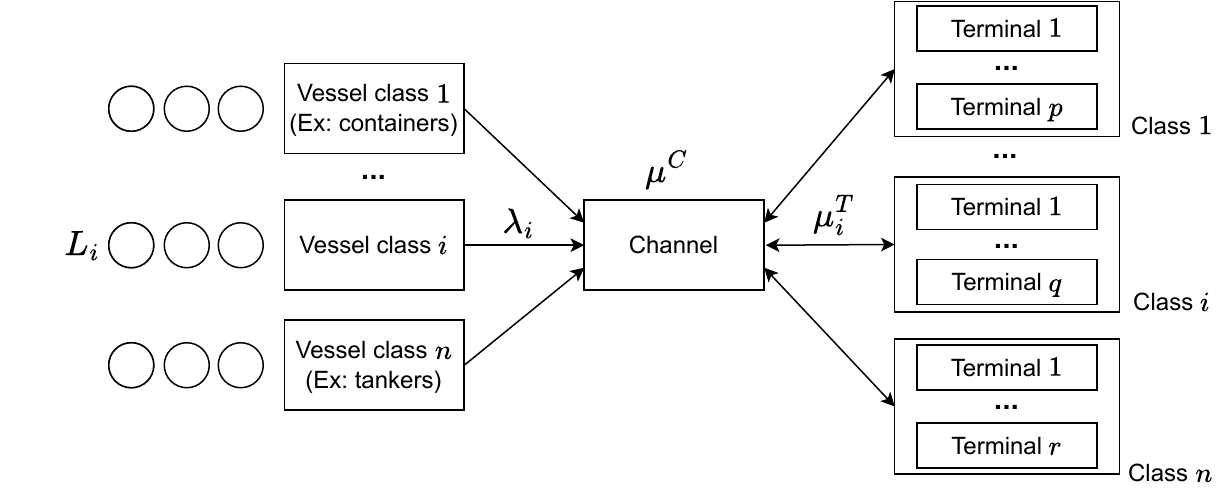}
    \caption{Structure of port queueing model.}
    \label{fig:queuestructure}
\end{figure}

We model the channel as a multiclass $G/M/1$ queue with equal priorities, and each terminal as an independent $G/M/1$ queue. Several studies, such as \citet{jones1968ship} and \citet{schonfeld1984optimizing} have adopted exponentially distributed service times for port subsystems.  Consistent with these studies, we assume that both the channel and terminal servers follow exponentially distributed service times. We estimate the service rate of channel and each terminal system from arrival distributions and observed queue lengths. The service rate represents the maximum rate at which a system can process vessels and hence is a measure of operating capacity of the system. The notations used in this study are summarized in Table~\ref{tab:notations}, and the queueing network structure is illustrated in Figure~\ref{fig:queuestructure}.

\begin{table}
\centering
\small
\begin{tabular}{c|l|l}
\hline
\textbf{Symbol} & \textbf{Meaning} & \textbf{Units} \\
\hline
$\tau$ & Time horizon & hours \\
\hline
$\lambda_i$ & Vessel arrival rate to the system (at anchorage) for class $i$ & vessels / hour\\
$c_i$ & Coefficient of variation for inter-arrival times of vessels (at anchorage) for class $i$ & -- \\
$\lambda$ & Aggregate vessel arrival rate to the system (at anchorage) & vessels / hour\\
$c_a$ & Coefficient of variation for inter-arrival times of vessels (at anchorage) & --\\
$L_i^C$ & Mean queue length of vessels of class $i$ at the anchorage due to channel constraints & vessels\\
$L_i^T$ & Mean queue length of vessels of class $i$ at the anchorage due to terminal constraints & vessels\\
$L_i$ & Mean queue length of vessels of class $i$ at the anchorage & vessels\\
$L$ & Mean queue length of vessels at the anchorage & vessels\\
$W^C$ & Mean waiting time at anchorage for all vessel classes due to channel constraints & hour \\
$W^T_i$ & Mean waiting time at anchorage for vessel class $i$ due to terminal constraints & hour \\
$\mu^C$ & Operating capacity of the channel & vessels / hour\\
$\mu^T_i$ & Operating capacity of terminal class $i$ & vessels / hour\\
$C_o$ & Operating capacity of the port & vessels / hour\\
$\rho^C$ & Utilization of channel & --\\
$\rho^T_i$ & Utilization of terminal class $i$ & --\\
\hline
$N^{\text{exits}}_\tau (\lambda)$ & Number of vessel exits from port in time $\tau$ when arrival rate is $\lambda$ & vessels / hour\\
$C_u$ & Ultimate capacity of port & vessels / hour\\
$C_s$ & Saturated capacity of port & vessels / hour\\
$\lambda_c$ &  Vessel arrival rate at ultimate capacity & vessels / hour\\
$\theta$ & Transition parameter of uncongested region & -- \\
$\beta$ & Transition parameter of congested region & -- \\
\hline
\end{tabular}
\caption{Notations, their meanings, and units.}
\label{tab:notations}
\end{table}

The expected waiting time due to channel constraints ($W^{C}$) for a multiclass $G/M/1$ queue with equal priorities can be approximated using Kingman’s formula \citep{kingman1961single}:
\begin{gather}
\mathbb{E}(W^{C}) \approx 
\left(\frac{\rho^{C}}{1-\rho^{C}}\right)
\left(\frac{1+c_a^2}{2}\right)
\left(\frac{1}{\mu^C}\right),
\label{channel:waittime}
\end{gather}
where $\rho^{C}$ is the utilization of the channel and $c_a$ is the coefficient of variation of inter-arrival times across all vessel classes. $\rho^{C}$ can be computed as $\rho^C = \lambda / \mu^C$, where $\mu^C$ is the operating capacity of the waterway channel. If arrivals are assumed follow a Poisson's process ($c_a = 1$), the approximation in equation~\eqref{channel:waittime} reduces to an exact expression for the expected waiting time.

An estimate of $\mathbb{E}(W^{C})$ can be obtained from vessel logs or directly observed from simulation results of port operations. Using a similar approach as \citet{lantz2017using}, equation~\eqref{channel:waittime} can be rearranged to estimate the operating capacity of the channel as
\begin{align}
    \mu^C = 
    \frac{W^C \lambda + 
    \sqrt{W^C \lambda \left(W^C \lambda + 2(c_a^2 + 1)\right)}}{2 W^C}. \label{eq:kingman}
\end{align}

Once the operating capacity of the channel ($\mu^C$) is determined, its utilization  can be computed as $\rho^C = \lambda / \mu^C$. Channel utilization provides an indication of how close the channel is to its operating capacity at the current arrival rate. Furthermore, using the expected waiting time due to channel constraints, we can estimate the queue length for each vessel class using Little’s law \citep{little2008little} as
\begin{equation}
\mathbb{E}(L^C_i) = \lambda_i \mathbb{E}(W^{C}). \label{eq:channel_queue}
\end{equation}
The observed queue lengths of different vessel classes at the anchorage can then be used to separate the contributions of channel and terminal constraints. Suppose $\mathbb{E}(L_i)$ denote the observed queue length at the anchorage for vessel class $i$, which can be obtained from AIS data or simulation outputs. Since $\mathbb{E}(L^C_i)$ is known from equation \eqref{eq:channel_queue}, the expected queue length attributable to terminal constraints can be estimated as
\begin{equation}
\mathbb{E}(L^T_i) = \mathbb{E}(L_i) - \mathbb{E}(L^C_i).
\end{equation}
Using Kingman’s formula along with Little’s law, the operating capacity of each terminal class can then be estimated as
\begin{equation}
    \mu^T_i = 
    \frac{\lambda_i}{2} + 
    \frac{\sqrt{\lambda_i^2 (L_i^T)^2 + 2 L_i^T \lambda_i^2 (c_i^2 + 1)}}{2 L_i^T},
\end{equation}
where $c_i$ is the coefficient of variation of inter-arrival times for vessels of class $i$. Similar to the channel, the utilization for each vessel class can be computed as $\rho^T_i = \lambda_i / \mu_i^T$. These utilization values provide insights into the congestion level of each vessel class within the port. 

Since vessels must pass through both the waterway channel and their respective terminal berths to be fully processed, the slower of these two subsystems determines the port’s effective throughput. If the channel can move vessels faster than the terminals can serve them, then terminal service becomes the limiting factor. Conversely, if the terminals collectively provide more service capacity than the channel can accommodate, the channel becomes the bottleneck. Accordingly, the overall operating capacity of the port is computed as the minimum of the operating capacities of the channel system and the downstream terminal system:
\begin{equation}
C_o = \min \left(\mu^C, \sum_i \mu_i^T \right).
\end{equation}

The estimated operating capacity of the port can be validated in two ways. First, given the operating capacities of individual terminals, class-specific anchorage waiting times $(W_i^{T})$ can be predicted using Kingman’s formula in \eqref{eq:kingman}. Comparing these predicted waiting times with AIS-observed waiting times provides a direct validation of the operating capacity estimates. Second, the operating capacity can also be independently inferred using a discrete-event simulation of the port system. As simulated arrival rates increase, the anchorage queue becomes unstable and grows without bound once the operating capacity is exceeded. The arrival rate at which this queueing behavior emerges provides a validation of the operating capacity estimate. In practice, however, identifying the precise arrival rate at which this transition occurs is challenging, as the shift from stable to unstable queueing behavior is gradual rather than abrupt. The analytical technique described in this section enables this transition point to be identified accurately.

\section{Ultimate capacity}\label{subsec:ul_cap}

The operating capacity model relies on the assumption of stable anchorage queue formation. Therefore, the model cannot be directly applied to estimate capacity under unstable conditions. During periods of instability, the queue length increases with each arriving vessel, as the system cannot process arrivals at the same rate. Consequently, the number of vessels processed becomes lower than the number of arrivals in a given time horizon. The state of the system when the number of exits from anchorage does not increase despite increasing arrivals defines the ultimate capacity of the port.

To estimate the ultimate capacity of the port, we first develop a detailed simulation of the full port system. The simulation is executed under a range of arrival rates $\lambda^{(i)}$, and for each run the number of vessels exiting the anchorage in given time horizon $\tau$, $\hat{N}^{\text{exits}}_\tau(\lambda^{(i)})$, is recorded. These data points are then fit to a theoretical ODE model that describes how the number of processed vessels changes with the arrival rate. The fitted parameters yield the port’s ultimate capacity. 

\subsection{Formulation \label{subsec:formulation}}

If all vessel classes are assumed to be identical, then at low arrival rates the number of vessels entering and leaving the anchorage is approximately equal, and any queues that form remain bounded. We denote the number of vessel arrivals at the port as $N^{\text{entries}}_\tau(\lambda)$ and the number of anchorage exist as $N^{\text{exits}}_\tau(\lambda)$ in a time horizon $\tau$. With a constant arrival rate $\lambda$, the number of arrivals observed over horizon $\tau$ is $N^{\text{entries}}_\tau(\lambda)=\lambda\tau$. In the uncongested regime, equal number of vessels arrive and exit, so $N^{\text{entries}}_\tau(\lambda)=N^{\text{exits}}_\tau(\lambda)$. As $\lambda$ increases, however, exits eventually plateau at the port’s ultimate capacity. This behavior is captured using the ODE
\begin{align}
\frac{dN_\tau^{\text{exits}}}{d\lambda}
= \tau \left(1 - \left(\frac{N_\tau^{\text{exits}}}{\tau C_u}\right)^{\theta}\right).
\label{eq:growth}
\end{align}

The ODE in equation \eqref{eq:growth} models the growth phase of the exit curve. At lower arrival rates, where $N_\tau^{\text{exits}}(\lambda) \ll \tau C_u$, the term $\left(1 - \frac{N_\tau^{\text{exits}}(\lambda)}{\tau C_u}\right) \approx 1$ in equation \eqref{eq:growth}, yielding $\frac{dN_\tau^{\text{exits}}}{d\lambda} \approx \tau$. This implies a linear relationship, such that $N_\tau^{\text{exits}}(\lambda) \approx \lambda \tau = N_\tau^{\text{entries}}(\lambda)$. At higher arrival rates, near ultimate capacity ($N_\tau^{\text{exits}}(\lambda) \approx \tau C_u$), the term $\left(1 - \frac{N^{\text{exits}}}{\tau C_u}\right) \approx 0$, leading to a flattening of the exit curve. The parameter $\theta$ controls how sharply the curve transitions from linear growth to the plateau, reflecting how quickly congestion manifests as arrivals increase. A higher value of $\theta$ indicates that the port has less ability to absorb rising arrivals, causing throughput to taper off more abruptly as the system approaches its ultimate capacity.

In multiclass systems, exits need not be monotone in $\lambda$ because vessel classes compete for shared resources such as pilots and channel capacity. The relationship between arrivals and exits in such cases is driven by two forces. The first force arises from terminal processing: vessels of classes with faster terminal service can move through the system more quickly compared to classes with slower terminal service. The second force arises from competition under first-come-first-served (FCFS) vessel dispatch. Once the shared resources become fully utilized, FCFS rule determines which vessel moves next. Under FCFS dispatch, the class with the older queues, that is, the classes whose vessels have waited longer due to their arrival pattern and terminal processing rates, is prioritized. Classes with younger queues are selected less frequently, and their throughput can decline even if the total arrival rate continues to increase. Throughput for each class therefore reflects the balance between these two forces.

The ODE in \eqref{eq:growth} must include a decline phase to capture this competitive effect. A logistic-type function is appropriate to model the decline phase because the decline phase causes exits to fall from one asymptote (the ultimate capacity) toward another asymptote (the saturated capacity). Although such extreme arrival rates are unlikely in practice, modeling this decline is important for fitting the simulated data and for correctly identifying the ultimate capacity. We therefore extend \eqref{eq:growth} to
\begin{align}
\frac{dN_\tau^{\text{exits}}}{d\lambda} =
\begin{cases}
\tau \left(1 - \left(\frac{N_\tau^{\text{exits}}}{\tau C_u}\right)^{\theta}\right), & \lambda \le \lambda_c, \\[6pt]
\frac{1}{\beta} (N_\tau^{\text{exits}} - \tau C_u) (N_\tau^{\text{exits}} - \tau C_s), & \lambda > \lambda_c.
\end{cases}
\label{eq:ode2}
\end{align}
Equation~\eqref{eq:ode2} assumes that exits initially increase with arrival rates in the same manner as in~\eqref{eq:growth}, until the shared resources reach their capacity. Beyond this point, increasing arrivals reduce the number of exits from the anchorage, marking the onset of congested regime. The parameter $\lambda_c$ denotes the arrival rate at which the port transitions into congested regime. This implies that the port attains its ultimate capacity, $C_u$, at an arrival rate of $\lambda_c$. Below $\lambda_c$, increases in arrivals lead to higher throughput, whereas beyond $\lambda_c$, further increases in arrivals reduce throughput due to increased competition between classes.

In equation \eqref{eq:ode2}, when $N^{\text{exits}}_\tau(\lambda)$ is close to ultimate capacity ($\tau C_u$), the term $(N_\tau^{\text{exits}} - \tau C_u) \approx 0$, allowing continuous shift from the growth phase to the decline region. Conversely, when $N^{\text{exits}}_\tau(\lambda)$ approaches saturated capacity ($\tau C_s$), the term $(N^{\text{exits}}_\tau - \tau C_s) \approx 0$ causes the throughput to level off at $C_s$. The parameter $\beta$ governs how steeply the system moves from the transition regime to the congested regime. A smaller $\beta$ indicates gradual transition into congested regime and a larger $\beta$ indicates a faster transition. In other words, a smaller $\beta$ means the higher processing rates can be sustained for a wider range of arrival rates. 

\subsection{Parameter estimation}

The ODE in \eqref{eq:ode2} does not admit an elementary closed-form solution. Consequently, the parameters of \eqref{eq:ode2} must be inferred numerically. This requires generating a set of arrival rates, ${\lambda^{(i)}}$, and corresponding throughput, ${\hat{N}^{\text{exits}}_\tau(\lambda^{(i)})}$ using a detailed port simulation model. Given $m$ such observations, the parameters of the ODE can be estimated by minimizing the mean squared error between the simulated exits and the ODE-predicted exits. The optimization problem in \eqref{eq:opti_obj}--\eqref{eq:opti_bounds} is then used to identify the optimal values of $C_u$, $C_s$, $\theta$, $\beta$, and $\lambda_c$.

\begin{align}
\min_{\boldsymbol{\eta} = (C_u, C_s, \theta, \beta, \lambda_c)} \quad &
\sum_{i=1}^{m} \left(N_\tau^{\text{exits}}(\lambda_i; \boldsymbol{\eta}) - \hat{N}^{\text{exits}}_\tau(\lambda^{(i)})\right)^2 
\label{eq:opti_obj} \\[10pt]
\text{subject to:} \quad &
\begin{aligned}
\frac{dN_\tau^{\text{exits}}}{d\lambda} &=
\begin{cases}
\tau \left(1 - \left(\dfrac{N_\tau^{\text{exits}}}{\tau C_u}\right)^{\theta}\right), & \lambda \le \lambda_c\\[6pt]
\frac{1}{\beta} \left(N_\tau^{\text{exits}} - \tau C_u\right)\left(N_\tau^{\text{exits}} - \tau C_s\right), & \lambda > \lambda_c
\end{cases}
\end{aligned}
\label{eq:opti_ode_constraint} \\[10pt]
& N_\tau^{\text{exits}}(0; \boldsymbol{\eta}) = 0 \label{eq:opti_ivp} \\[6pt]
& C_u, C_s, \lambda_c, \theta, \beta \ge 0 \label{eq:opti_bounds}
\end{align}\

We denote the decision variables $(C_u, C_s, \theta, \beta, \lambda_c)$ as a candidate vector $\boldsymbol{\eta}$. We use $N_\tau^{\text{exits}}(\lambda;\boldsymbol{\eta})$ to denote the solution of the ODE in~\eqref{eq:ode2} evaluated at arrival rate $\lambda$ using parameter vector $\boldsymbol{\eta}$. The objective in \eqref{eq:opti_obj} minimizes the mean squared error (MSE) between the ODE-predicted exits and the simulated exits for each simulation run $i = 1, \dots, m$ over varying arrival rates. Constraints \eqref{eq:opti_ode_constraint} and \eqref{eq:opti_ivp} ensure that the predicted exits satisfy the ODE in \eqref{eq:ode2} with the initial condition $N_\tau^{\text{exits}}(0; \boldsymbol{\eta})=0$, which states that no vessels can exit when there are no arrivals. Finally, constraint \eqref{eq:opti_bounds} ensures non-negativity for the decision variables. The optimization problem in \eqref{eq:opti_obj}--\eqref{eq:opti_bounds} is nonlinear and includes a differential equation constraint, but it contains only a small number of decision variables. We therefore solve it using a differential evolution algorithm \citep{storn1997differential}. Differential evolution maintains a population of candidate vectors and improves them through mutation and crossover. While using differential evolution, restricting variables to reasonable ranges helps reduce the search space and improves convergence. In practice, engineering judgment allows us to set upper bounds on $\lambda_c$, $C_u$, and $C_s$, often taken as multiples of the current port arrival rate. Parameters $\theta$ and $\beta$ perform well within the ranges $0$--$100$ and $0$--$1000$, respectively. A high-level summary of the process is given in Algorithm \ref{alg:integrated_de_fit}.

In line 1 of Algorithm \ref{alg:integrated_de_fit}, an initial population of such vectors is generated. For each vector, the ODE in \eqref{eq:ode2} is solved numerically as an initial value problem, and its fitness is computed as the sum of squared errors (lines 2--3). The loop in lines 4--16 performs the differential evolution updates for $MG$ generations. In each generation, for every population vector $\boldsymbol{\eta}_j$, three distinct vectors are randomly selected to construct a mutant vector, $\boldsymbol{v}_j$ (lines 6--7). The mutant vector, $\boldsymbol{v}_j$, is then combined with $\boldsymbol{\eta}_j$ using binomial crossover to produce a trial vector $\boldsymbol{u}_j$ (line 8).
 The trial vector is projected back into its parameter bounds if necessary (line 9), after which the ODE is solved again and its fitness is computed (lines 10--11). If the trial vector achieves a lower MSE than the current vector, it replaces it (lines 12--13). The procedure ends when all generations are processed or MSE is zero for any vector (line 14) at any point. The best-performing vector in the final population is returned as the optimal solution (lines 16--17).

 \begin{algorithm}
\caption{ODE-constrained parameter fitting via differential evolution}
\label{alg:integrated_de_fit}
\begin{algorithmic}[1]
\Require 
  Observed arrival rates $\{\lambda^{(i)}\}_{i=1}^m$, simulated exits $\{\hat{N}^{\text{exits}}_\tau(\lambda^{(i)})\}_{i=1}^m$, time horizon $\tau$, parameter bounds $\boldsymbol{\eta}_{\text{bounds}}$, population size $P$, mutation factor $F$, crossover rate $CR$, maximum generations $MG$.
\Ensure 
  Optimal parameter vector $\boldsymbol{\eta}^* = (C_u^*, C_s^*, \theta^*, \beta^*, \lambda_c^*)$.
  
\State Initialize a population of $P$ candidate vectors $\{\boldsymbol{\eta}_1,\dots,\boldsymbol{\eta}_P\}$ by sampling within given bounds.
\For{$j = 1$ to $P$}
  \State Compute fitness $E_j = \displaystyle \sum_{i=1}^n \left(N_\tau^{\text{exits}}(\lambda^{(i)};\boldsymbol{\eta}_j) - \hat{N}^{\text{exits}}_\tau(\lambda^{(i)})\right)^2$, 
  where $N_\tau^{\text{exits}}(\cdot;\boldsymbol{\eta}_j)$ is obtained by numerically solving the ODE in \eqref{eq:ode2} with $N_\tau^{\text{exits}}(0)=0$ with parameters $\boldsymbol{\eta}_j$.
\EndFor

\For{generation $= 1$ to $MG$}
  \For{$j = 1$ to $P$}
    \State Select three distinct indices $r_1, r_2, r_3 \in \{1,\dots,P\}$ with $r_1 \neq r_2 \neq r_3 \neq j$.
    \State Mutation: form mutant vector $\boldsymbol{v}_j = \boldsymbol{\eta}_{r_1} + F \cdot (\boldsymbol{\eta}_{r_2} - \boldsymbol{\eta}_{r_3})$.
    \State Crossover: form trial vector $\boldsymbol{u}_j$ by combining components of $\boldsymbol{v}_j$ and $\boldsymbol{\eta}_j$ using crossover rate $CR$.
    \State Project $\boldsymbol{u}_j$ back into $\boldsymbol{\eta}_{\text{bounds}}$ if any component is outside its bounds.
    \State Numerically integrate the ODE in equation \eqref{eq:ode2} and record $N_\tau^{\text{exits}}(\lambda^{(i)};\boldsymbol{u}_j)$ at each $\lambda^{(i)}$.
      \State Compute
      \[
      E_{\text{trial}} = \sum_{i=1}^m \left(N_\tau^{\text{exits}}(\lambda^{(i)};\boldsymbol{u}_j) - \hat{N}^{\text{exits}}_\tau(\lambda^{(i)})\right)^2.
      \]
    
    \If{$E_{\text{trial}} < E_j$}
      \State Set $\boldsymbol{\eta}_j \gets \boldsymbol{u}_j$ and $E_j \gets E_{\text{trial}}$.
    \EndIf
  \EndFor
  
  \If{$\min_{j}E_j = 0$}
    \State break
  \EndIf
\EndFor

\State Find $\boldsymbol{\eta}^* = \boldsymbol{\eta}_{\arg\min E_j}$ in the final population.
\State \Return $\boldsymbol{\eta}^*$
\end{algorithmic}
\end{algorithm}

\subsection{Relation between operating and ultimate capacity}\label{subsec:rel}

The operating capacity and ultimate capacity are estimated using two different approaches. The operating capacity model discussed in Section \ref{subsec:op_cap} relies on a queueing formulation, while the ultimate capacity model presented in Section \ref{subsec:formulation} uses data from a discrete-event simulation of the port fitted to an ODE model. However, the ODE formulation in \eqref{eq:ode2} is also consistent with the definition of operating capacity. Under stable conditions, operating capacity assumes that entries and exits are equal over the horizon $\tau$, so $N_\tau^{\text{exits}} = \lambda \tau$ and $dN_\tau^{\text{exits}}/d\lambda = \tau$. At low arrival rates, the ODE model reproduces this behavior naturally. When $N_\tau^{\text{exits}} \ll \tau C_u$, the congestion term $\left(1 - \left(\frac{N_\tau^{\text{exits}}}{\tau C_u}\right)^{\theta}\right)$ is close to one, yielding $dN_\tau^{\text{exits}}/d\lambda \approx \tau$, which corresponds to stable operations.

To connect this with operating capacity more formally, consider the ODE
\begin{align}
\frac{dN_\tau^{\text{exits}}}{d\lambda}
= \tau\!\left(1 - \left(\frac{N_\tau^{\text{exits}}}{\tau C_u}\right)^\theta\right),
\qquad N_\tau^{\text{exits}}(0)=0.
\end{align}

When $\theta$ is large, the term $\left(N_\tau^{\text{exits}}/(\tau C_u)\right)^{\theta}$ remains negligible for all $N_\tau^{\text{exits}} < \tau C_u$, so $dN_\tau^{\text{exits}}/d\lambda \approx \tau$ for all $\lambda$. Only when $N_\tau^{\text{exits}}$ approaches $\tau C_u$ does the derivative rapidly collapse toward zero. Thus, increasing $\theta$ sharpens the transition between stable and saturated behavior. In the limit as $\theta \to \infty$, the slope remains equal to $\tau$ until $N_\tau^{\text{exits}}$ is arbitrarily close to $\tau C_u$. In this regime, $N_\tau^{\text{entries}} = N_\tau^{\text{exits}}$, and the system remains uncongested. Consequently, the operating and ultimate capacities become arbitrarily close as $\theta \to \infty$.

Further, at operating capacity, the stability condition requires $N_\tau^{\text{exits}} = C_o \tau$. Integrating the ODE from arrival rate $0$ (where $N_\tau^{\text{exits}}=0$) to arrival rate $C_o$ (where $N_\tau^{\text{exits}}=C_o \tau$) yields
\begin{align}
\int_0^{C_o \tau}
\frac{1}{\tau\left(1 - \left(N_\tau^{\text{exits}}/(\tau C_u)\right)^\theta\right)}\, dN_\tau^{\text{exits}}
= C_o.
\label{eq:direct_integral_tau_subscript}
\end{align}

The integral in \eqref{eq:direct_integral_tau_subscript} is finite only when $C_o < C_u$. To see this, observe that the integrand is well defined as long as $N_\tau^{\text{exits}} < \tau C_u$, since the denominator
$1 - (N_\tau^{\text{exits}}/(\tau C_u))^\theta$ remains non-zero.
As $N_\tau^{\text{exits}}$ approaches $\tau C_u$, the denominator approaches zero, causing the integrand to diverge. Consequently, the integral converges only if the upper limit satisfies $C_o \tau < \tau C_u$, or equivalently $C_o < C_u$.
As $C_o \to C_u$, the integral becomes unbounded. This establishes that the
operating capacity must always satisfy $C_o < C_u$. In summary, we find that:
\begin{itemize}
\item When arrivals are low, both operating capacity and ultimate capacity formulations yield similar estimates of throughput. When $N_\tau^{\text{exits}} \ll \tau C_u$, the ODE simplifies to $dN_\tau^{\text{exits}}/d\lambda \approx \tau$, matching the linear, stable uncongested regime assumed in the operating-capacity definition.

\item In the limit $\theta \to \infty$, exits remain arbitrarily close to arrivals until the ultimate capacity is reached, after which throughput becomes constant. Increasing $\theta$ sharpens the transition between stable and saturated regimes. When $\theta \to \infty$, the operating and ultimate capacities become arbitrarily close. 

\item Operating capacity is always strictly below ultimate capacity for any finite $\theta$. Practically, this means ports generally possess some limited ability to process arrivals above the long-run operating capacity. However, these conditions are unstable and would cause significant congestion for later arriving vessels. 
\end{itemize}

\section{Computational results}\label{sec:results} 

We tested the capacity models presented in this study using a data-driven discrete event simulation of the Port of Houston. The model represents the Houston Ship Channel, 27 aggregated ship-to-shore terminals, and associated landside connections via truck, rail, and pipelines. These terminals handle containerized, liquid-bulk, and non-containerized cargo (e.g., dry-bulk, break-bulk, general cargo). The Port of Houston is one of the largest in the US, ranking first in foreign waterborne tonnage and fifth in total TEUs among US container ports~\citep{PortHoustonStats}. The high complexities and interdependencies make the Port of Houston an ideal candidate for validating the models proposed in this study. In this section, we briefly describe the simulation development
in Section~\ref{subsec:res_sim}, present results on operating capacity
in Section~\ref{sec:op_cap_results}, and discuss ultimate capacity results
in Section~\ref{subsec:res_ul_cap}.

All experiments in this study were conducted on a Dell Precision 3680 workstation and all optimization problems were solved using the SciPy package \citep{2020SciPy-NMeth} in Python 3.12. ODEs were solved numerically using the \texttt{integrate} module from the same package.

\subsection{Simulation development}\label{subsec:res_sim}

We modeled the port system processes using a discrete-event simulation, integrating three interlinked subsystems: waterside (anchorage and channel), terminals (container, tanker, and non-containerized cargo), and landside (trucking, rail, and pipeline) operations. The decision rules covered anchorage queueing, vessel assignment to terminals, vessel scheduling under navigational restrictions, berth allocation, cargo-handling processes at terminals, and landside activities. While the full simulation structure is too extensive to detail here, we summarize the core processes and present validation results supported by real-world data comparisons. A complete description of the simulation model, including calibration and validation procedures and results, is provided in \citealp{bagchi2025dynamic}. The main components of the simulation are as follows:
\begin{itemize}
    \item \textbf{Anchorage and waterway channel system:} Vessels arrive at the anchorage according to a Poisson process with arrival rates derived from archival AIS records. These vessels remain queued until berths are available at their assigned terminals. When reaching the front of the queue, a vessel requests channel entry, governed by a scheduling algorithm that ensures compliance with navigational, spacing, and daylight restrictions. These restrictions cause differential anchorage waiting to different vessels based on their dimensions and weight. Entry to the channel is permitted only when berth, pilot, tug, and channel restrictions allow the vessel to pass through the entire channel without waiting midway. The channel is discretized into segments defined by width, depth, and spacing constraints, with a dockline-based \citep{zohoori2022quantifying} scheduling algorithm to model this behavior. 

    \item \textbf{Terminal systems:} Each terminal incorporates process models for cargo handling (loading, unloading, yard transfer), resource allocation (cranes, pumps, conveyors), and integration with landside operations. The cargo weight and type (containerized, liquid-bulk, non-containerized cargo) of different vessels along with the characteristic of their destination terminals such as storage capacities, berth counts, and transfer rates determine the dwell times of vessels. In addition to physical resource limitations, terminal efficiencies are also modeled to accurately represent vessel dwell times. 
    
    \item \textbf{Landside subsystem:} The landside model incorporates trucking, rail, and pipeline infrastructure, mapped to each terminals. Truck arrivals follow a uniform arrival rate between gate timings, and a truck queueing model determine truck turn times. Rail arrival rates and car counts determine lead times for cargo departures on rail. Pipelines, when applicable, operate as variable-rate inflow or outflow processes triggered by storage-level thresholds. These subsystems influence terminal utilization by controlling cargo movement between storage areas and external networks. 
\end{itemize}

The model was calibrated using real-world data from Port of Houston. The model validation was then performed in three stages to ensure applicability of the model. First, we determined the number of replicate simulation runs required to obtain stable outputs, given that the model included several stochastic parameters. Confidence intervals for anchorage queue length means were computed using \textit{t}-statistics. A 95\% two-tailed confidence intervals were created for mean queue length outputs from different number of replications of simulation. The mean anchorage queue length and the width of 95\% confidence interval for the mean is shown in Table \ref{tab:queue_ci}. It can be observed the mean and standard deviation of the anchorage queue length remain consistent for simulations with five or more replications. These results indicate that the minimum number of runs needed for consistent numerical output is five. Hence, for the rest of the analyses conducted in this study use average values of six replicate runs. 
\begin{table}
\centering
\begin{tabular}{ccccc}
\hline
\textbf{Number of replicate runs} & \textbf{Mean} & \textbf{Stdev} & \textbf{95\% CI} & \textbf{95\% CI width} \\
\hline
3  & 27.3 & 4.8 &  (15.38, 39.22) & 23.84 \\
5  & 26.4 & 5.3 & (19.82, 32.98) & 13.16 \\
8  & 26.6 & 5.7 & (21.83, 31.37) & 9.54  \\
10 & 26.4 & 5.6 &  (22.39, 30.41) & 8.02  \\
12 & 26.6 & 5.5 &  (23.11, 30.09) & 6.98  \\
15 & 26.4 & 5.5 & (23.35, 29.45) & 6.10  \\
\hline
\end{tabular}
\caption{Simulation queue-length confidence-interval analysis. (Mean: mean anchorage queue length;
Stdev: standard deviation;
95\% CI: 95\% confidence interval for the mean;
95\% CI width: width of the 95\% confidence interval.)\label{tab:queue_ci}}
\end{table}

\begin{figure}
    \centering
    \begin{subfigure}[b]{0.48\textwidth}
        \centering
        \includegraphics[width=\textwidth]{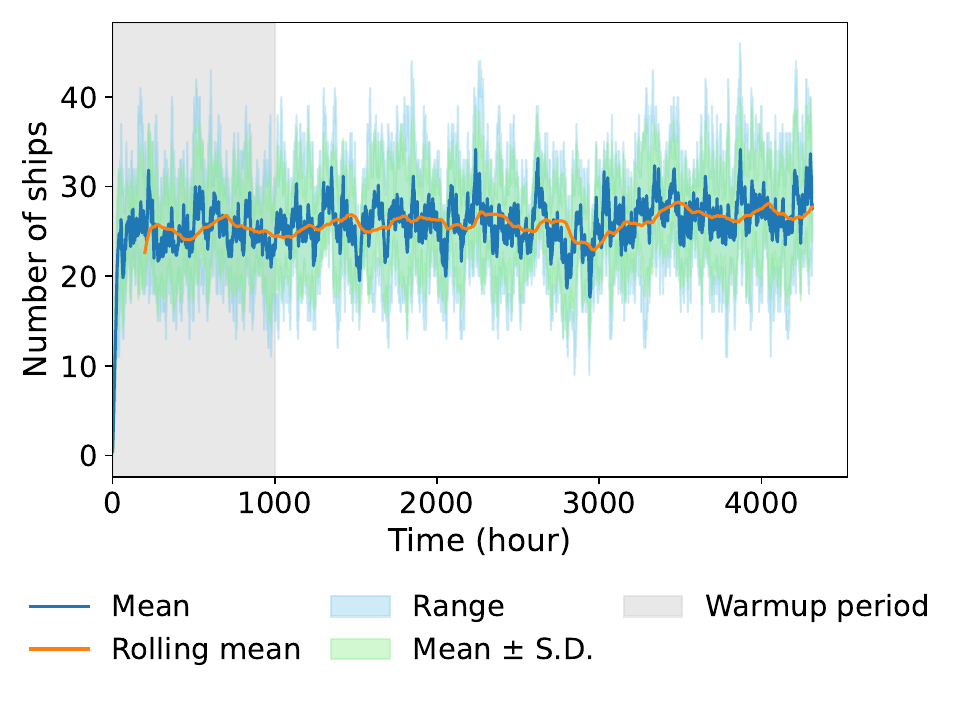}
        \caption{Anchorage queue.}
        \label{fig:sub1}
    \end{subfigure}
    \hfill
    \begin{subfigure}[b]{0.48\textwidth}
        \centering
        \includegraphics[width=\textwidth]{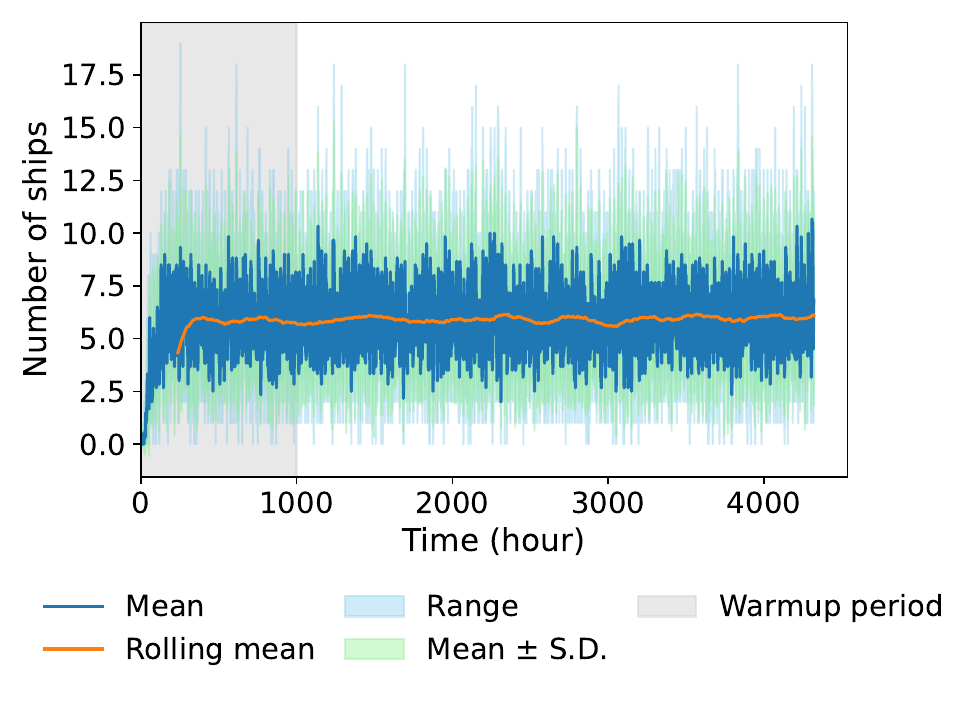}
        \caption{Channel occupancy.}
        \label{fig:sub2}
    \end{subfigure}
    \caption{Anchorage queue dynamics and channel occupancy observed over a six-month simulation horizon, based on six replicate runs.}
    \label{fig:anch_queue}
\end{figure}

Second, a temporal validation was performed to obtain a warmup period and ensure long term stability. In our study, we found a warm up period corresponding to the first 1000 hours to be sufficient to observe stable queueing behavior. Figure~\ref{fig:anch_queue} illustrates the anchorage queue dynamics and channel occupancy for a six-month simulation run, with the warm-up period indicated. Finally, outputs from a six-month simulation with six replicate runs were validated against AIS-derived observations of anchorage queue size and vessel waiting time. Table \ref{tab:sim_validation} summarizes the validation results.
Overall, the simulation shows good alignment with observed values across the three key validation metrics.

\begin{table}
\centering
\begin{tabular}{lcccccc}
\hline
\multirow{2}{*}{\textbf{Quantity}} & 
\multicolumn{3}{c}{\textbf{AIS values}} & 
\multicolumn{3}{c}{\textbf{Simulation values}} \\
\cmidrule{2-7}
 & \textbf{Container} & \textbf{Non-container} & \textbf{Tanker} & \textbf{Container} & \textbf{Non-container} & \textbf{Tanker} \\
\hline
Anchorage queue size     & 1.2  & 4.0  & 23.6 & 1.2  & 3.0  & 22.0 \\
Anchorage wait time (hr) & 14.8 & 27.8 & 50.7 & 16.6 & 36.9 & 70.7 \\
Terminal dwell time (hr) & 35.0 & 60.0 & 45.0 & 34.7 & 66.4 & 55.4\\
\hline
\end{tabular}
\caption{Validation of simulation results against AIS-derived values.}
\label{tab:sim_validation}
\end{table}

\subsection{Results on operating capacity}\label{sec:op_cap_results}

To evaluate the operating capacity model described in Section~\ref{subsec:op_cap} for the Port of Houston, archival AIS data from 2022 through 2024 were analyzed at a quarterly resolution. These data were used to estimate anchorage queue lengths and vessel waiting times by cargo class. The proportion of total anchorage waiting time attributable specifically to channel restrictions was obtained from the developed simulation model. It should be noted that estimation of operating capacity does not require a simulation when reliable estimates of channel-restriction waiting times are available from port operational logs directly.

The observed mean waiting time due to channel restrictions was $42.1$ minutes. This estimate is consistent with \cite{zohoori2022quantifying}, who reported $320.28$ hours of total channel-related delays over one month. Based on an arrival rate of approximately $0.7$ vph \citep{bathgateAIS} during that period, the implied average wait is $38$ minutes per vessel, which aligns closely with our estimated $42.1$ minutes. Using a mean waiting time of $42.1$~minutes due to channel restrictions, the channel operating capacity was computed to be $1.52$~vph. Table~\ref{tab:queue_analysis} reports AIS-derived vessel arrival rates, anchorage waiting times, anchorage queue lengths due to terminal restrictions, and the resulting operating capacity estimates.

The computed operating capacity was found to be approximately $0.82$--$0.98$ vessels processed by the port per hour with an increasing trend over the years. Comparing the port operating capacity with the channel operating capacity of $1.52$~vph, we find that channel restrictions are unlikely to be a capacity bottleneck at the Port of Houston under current operating conditions. It should be noted that this operating capacity estimate does not reflect the maximum number of vessels that the port can accommodate at any given time. Instead, it indicates the mean inbound number of vessels that the port can handle over an extended period. For example, the computed $C_o$ values in Table \ref{tab:queue_analysis} represent the maximum time-averaged hourly processing rate of inbound vessels over each three-month period. The operating capacity value accounts for all operating conditions including nights, weekends, holidays, channel restrictions, and terminal-level resource constraints.

Operating capacity ranged from $0.10$--$0.17$~vph for container vessels,
$0.17$--$0.21$~vph for non-container vessels, and $0.52$--$0.62$~vph for tanker vessels. The utilization, $\rho_i^T$, measures the level of congestion at each terminal class and ranged from $0.53$--$0.90$ for container terminals,
$0.75$--$0.88$ for non-container terminals, and $0.95$--$0.97$ for tanker terminals. The consistently higher utilization for tanker vessels indicate that tanker terminals operate closest to operating capacity at the Port of Houston, implying that even small increases in tanker arrivals can lead to disproportionately large increases in anchorage queue lengths. Utilization for container and non-containerized cargo terminals are lower in the later years, indicating greater slack available in these terminal classes. The exceptionally high  wait times and high utilization observed for container terminals in 2022 can be attributed to congestion effects associated with post-COVID-19 related disruptions.

We validated the derived operating capacity by comparing predicted anchorage wait times with AIS-derived wait times. The maximum relative error was $1.24\%$ across all vessel classes and quarters, reinforcing confidence in the capacity estimates. In addition to validation against AIS-derived observations, we validated the operating capacity model using simulation outputs alone. For each terminal class, arrival rates, queue lengths, and mean waiting times were extracted from the simulation for container, liquid-bulk, and non-containerized cargo terminals. The estimated operating capacity for container terminals was $0.15$~vph with a utilization of $0.64$. Tanker terminals exhibited an operating capacity of $0.57$~vph with a utilization of $0.96$, while non-containerized cargo terminals had an operating capacity of $0.18$~vph with a utilization of $0.79$.

Since congestion at the channel was minimal, terminal operations constitute the binding bottleneck for operating capacity at the Port of Houston. Based on the simulation results, the calculated operating capacity of the terminal system was $0.9$~vph, which also determines the operating capacity of the port. These results closely match the operating capacities and utilization derived from AIS data alone, as reported in Table~\ref{tab:queue_analysis}.

\begin{table}
  \centering
  \begin{tabular}{lllrrrrrrrrr}
    \hline
Year & Quarter & Vessel type & $\lambda_i$ & $c_i$ & $L_i^T$ & $\mu_i^T$ & $C_o$ & Calculated & Actual & Relative & $\rho_i^T$ \\
& & & & & & & & $W_i^T$ & $W_i^T$ & error (\%) & \\
    \cmidrule(lr){1-12}
    \multirow{12}{*}{2022} & \multirow{3}{*}{Q1} & Container & 0.09 & 1.0 & 6.0 & 0.100 & \multirow{3}{*}{0.820} & 70.1 & 69.9 & 0.37 & 0.872 \\
     &  & Non-container & 0.17 & 1.1 & 6.9 & 0.200 &  & 39.8 & 39.7 & 0.32 & 0.881 \\
     &  & Tanker & 0.49 & 1.1 & 20.2 & 0.520 &  & 41.3 & 41.2 & 0.09 & 0.950 \\
    \cmidrule(lr){2-12}
     & \multirow{3}{*}{Q2} & Container & 0.09 & 1.1 & 7.1 & 0.100 & \multirow{3}{*}{0.870} & 83.2 & 82.8 & 0.53 & 0.883 \\
     &  & Non-container & 0.17 & 1.1 & 4.3 & 0.210 &  & 24.7 & 24.6 & 0.31 & 0.824 \\
     &  & Tanker & 0.53 & 1.2 & 23.7 & 0.560 &  & 44.4 & 44.4 & 0.00 & 0.955 \\
    \cmidrule(lr){2-12}
     & \multirow{3}{*}{Q3} & Container & 0.10 & 1.1 & 9.4 & 0.110 & \multirow{3}{*}{0.840} & 99.1 & 99.1 & 0.08 & 0.904 \\
     &  & Non-container & 0.16 & 1.2 & 5.3 & 0.190 &  & 33.5 & 33.4 & 0.36 & 0.843 \\
     &  & Tanker & 0.52 & 1.2 & 22.7 & 0.540 &  & 43.8 & 43.7 & 0.11 & 0.953 \\
    \cmidrule(lr){2-12}
     & \multirow{3}{*}{Q4} & Container & 0.10 & 1.0 & 5.8 & 0.110 & \multirow{3}{*}{0.850} & 58.4 & 58.2 & 0.30 & 0.869 \\
     &  & Non-container & 0.17 & 1.0 & 5.7 & 0.200 &  & 33.3 & 33.3 & 0.07 & 0.868 \\
     &  & Tanker & 0.52 & 1.0 & 23.9 & 0.540 &  & 46.0 & 46.2 & 0.40 & 0.961 \\
    \cmidrule(lr){1-12}
    \multirow{12}{*}{2023} & \multirow{3}{*}{Q1} & Container & 0.09 & 1.0 & 3.0 & 0.120 & \multirow{3}{*}{0.860} & 31.4 & 31.2 & 0.58 & 0.784 \\
     &  & Non-container & 0.15 & 1.0 & 4.1 & 0.180 &  & 28.1 & 27.9 & 0.41 & 0.833 \\
     &  & Tanker & 0.54 & 1.1 & 28.1 & 0.560 &  & 51.9 & 51.8 & 0.00 & 0.964 \\
    \cmidrule(lr){2-12}
     & \multirow{3}{*}{Q2} & Container & 0.10 & 1.0 & 1.1 & 0.150 & \multirow{3}{*}{0.880} & 11.5 & 11.4 & 0.83 & 0.638 \\
     &  & Non-container & 0.15 & 1.0 & 3.8 & 0.180 &  & 25.4 & 25.3 & 0.43 & 0.824 \\
     &  & Tanker & 0.52 & 1.1 & 20.4 & 0.550 &  & 39.0 & 39.0 & 0.08 & 0.952 \\
    \cmidrule(lr){2-12}
     & \multirow{3}{*}{Q3} & Container & 0.10 & 1.0 & 2.1 & 0.130 & \multirow{3}{*}{0.870} & 21.2 & 21.1 & 0.62 & 0.745 \\
     &  & Non-container & 0.14 & 0.9 & 3.8 & 0.170 &  & 27.9 & 27.8 & 0.40 & 0.829 \\
     &  & Tanker & 0.55 & 1.1 & 26.2 & 0.570 &  & 47.4 & 47.4 & 0.04 & 0.962 \\
    \cmidrule(lr){2-12}
     & \multirow{3}{*}{Q4} & Container & 0.09 & 1.0 & 1.3 & 0.140 & \multirow{3}{*}{0.900} & 13.5 & 13.4 & 0.21 & 0.648 \\
     &  & Non-container & 0.13 & 1.0 & 2.4 & 0.170 &  & 18.0 & 17.9 & 0.45 & 0.760 \\
     &  & Tanker & 0.57 & 1.0 & 27.0 & 0.590 &  & 47.5 & 47.4 & 0.11 & 0.964 \\
    \cmidrule(lr){1-12}
    \multirow{12}{*}{2024} & \multirow{3}{*}{Q1} & Container & 0.10 & 1.0 & 1.8 & 0.140 & \multirow{3}{*}{0.910} & 18.3 & 18.2 & 0.60 & 0.720 \\
     &  & Non-container & 0.14 & 1.0 & 4.0 & 0.170 &  & 27.8 & 27.7 & 0.42 & 0.830 \\
     &  & Tanker & 0.58 & 1.1 & 26.0 & 0.600 &  & 44.9 & 44.9 & 0.15 & 0.962 \\
    \cmidrule(lr){2-12}
     & \multirow{3}{*}{Q2} & Container & 0.09 & 1.0 & 0.6 & 0.170 & \multirow{3}{*}{0.980} & 6.7 & 6.7 & 0.38 & 0.541 \\
     &  & Non-container & 0.14 & 1.1 & 2.5 & 0.190 &  & 17.4 & 17.4 & 0.09 & 0.754 \\
     &  & Tanker & 0.59 & 1.2 & 32.7 & 0.620 &  & 55.0 & 55.0 & 0.03 & 0.966 \\
    \cmidrule(lr){2-12}
     & \multirow{3}{*}{Q3} & Container & 0.09 & 0.9 & 0.6 & 0.170 & \multirow{3}{*}{0.940} & 6.1 & 6.1 & 1.24 & 0.525 \\
     &  & Non-container & 0.15 & 1.1 & 4.1 & 0.180 &  & 27.0 & 27.3 & 0.86 & 0.824 \\
     &  & Tanker & 0.57 & 1.2 & 35.3 & 0.590 &  & 61.7 & 61.9 & 0.28 & 0.967 \\
    \cmidrule(lr){2-12}
     & \multirow{3}{*}{Q4} & Container & 0.09 & 0.9 & 0.9 & 0.150 & \multirow{3}{*}{0.960} & 9.8 & 9.8 & 0.59 & 0.609 \\
     &  & Non-container & 0.16 & 1.0 & 3.5 & 0.200 &  & 21.4 & 21.4 & 0.06 & 0.816 \\
     &  & Tanker & 0.58 & 1.1 & 29.0 & 0.610 &  & 49.7 & 49.8 & 0.19 & 0.965 \\
    \hline
  \end{tabular}
  \caption{Operating capacity results. ($\lambda_i$: arrival rate for cargo class $i$ (vph); 
$c_i$: Coefficient of variation for inter-arrival time for cargo class $i$ ; 
$L_i^T$: mean terminal queue length (vessels); 
$\mu_i^T$: capacity for terminal class $i$ (vph); 
$C_o$: operating capacity of port (vph); 
Calculated $W_i^T$: model-predicted wait time at terminal class $i$ (hour); 
Actual $W_i^T$: AIS-observed wait time at terminal class $i$ (hour); 
$\rho_i^T$: terminal utilization.)}\label{tab:queue_analysis}
\end{table}

\subsection{Results on ultimate capacity}
\label{subsec:res_ul_cap}

\begin{figure}
    \centering
    \begin{subfigure}[]{0.48\textwidth}
        \centering
        \includegraphics[width=\textwidth, trim=0 0 0 0, clip]{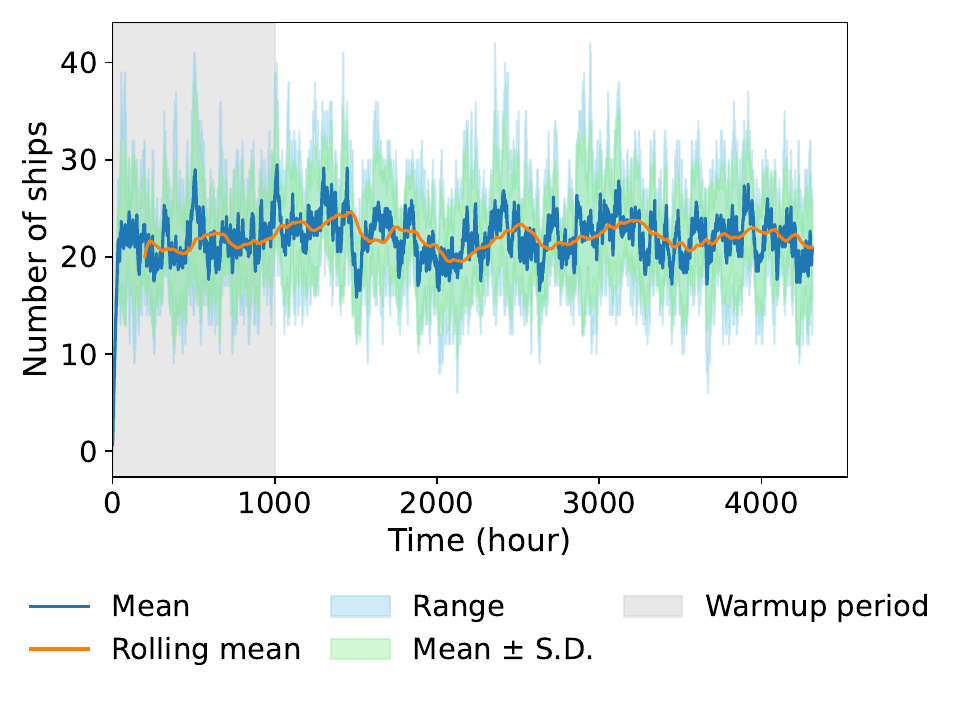}
        \caption{\centering $\lambda=0.7$~vph, $\lambda < C_o = 0.9$~vph, $L$ is bounded with $\mathbb{E}(L)  = 22.1$ (2968 vessels exited anchorage).}
    \end{subfigure}
    \hfill
    \begin{subfigure}[]{0.48\textwidth}
        \centering
        \includegraphics[width=\textwidth, trim=0 0 0 0, clip]{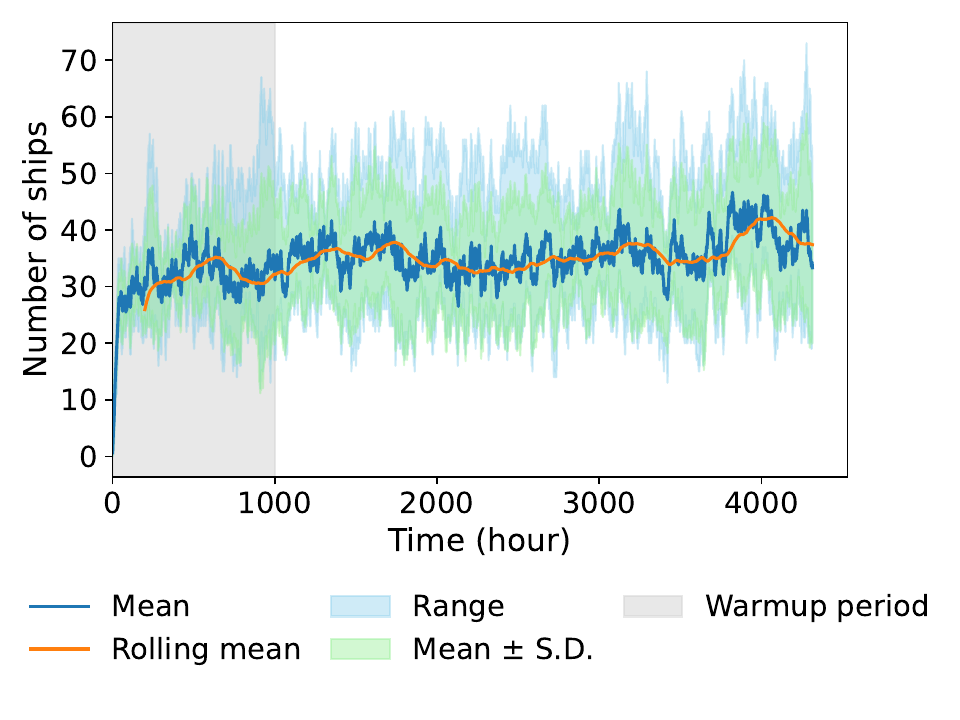}
        \caption{\centering $\lambda=0.9$~vph, $\lambda = C_o = 0.9$~vph, $L$ is bounded with $\mathbb{E}(L)  = 35.8$, (3832 vessels exited anchorage).}
    \end{subfigure}
    
    \vspace{0.5cm} 
    
    \begin{subfigure}[]{0.48\textwidth}
        \centering
        \includegraphics[width=\textwidth, trim=0 0 0 0, clip]{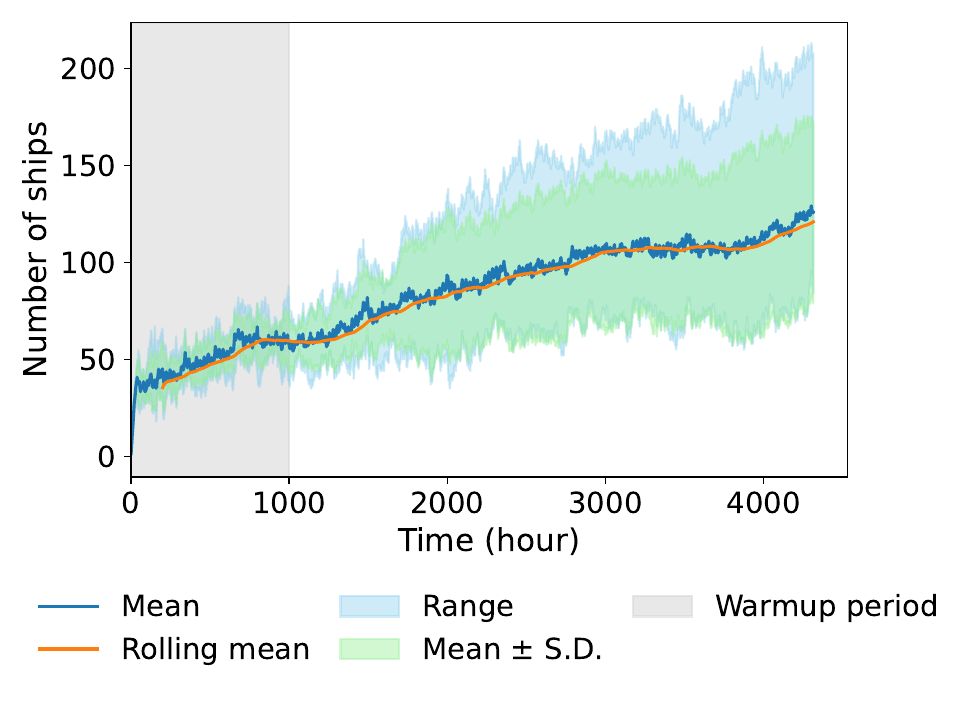}
        \caption{\centering $\lambda=1.1$~vph, $\lambda > C_o = 0.9$~vph, $L$ is unbounded, (4620 vessels exited anchorage).}
    \end{subfigure}
    \hfill
    \begin{subfigure}[]{0.48\textwidth}
        \centering
        \includegraphics[width=\textwidth, trim=0 0 0 0, clip]{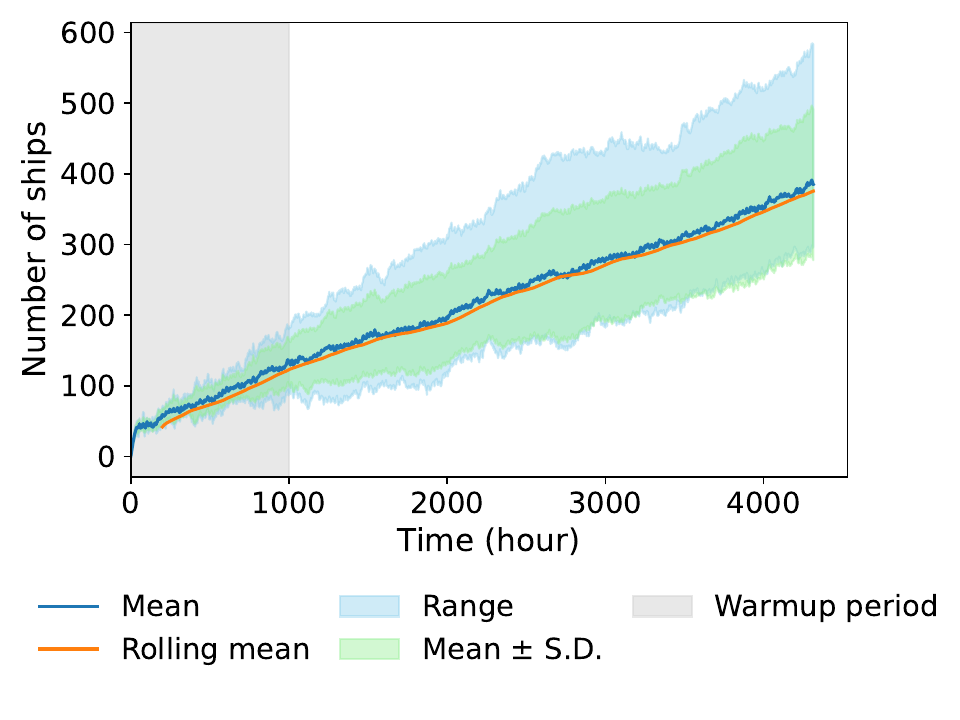}
        \caption{\centering $\lambda=1.3$~vph, $\lambda > C_o = 0.9$~vph, $L$ is unbounded, (5235 vessels exited anchorage).}
    \end{subfigure}
    
\caption{Anchorage queue behavior at different arrival rates over a six-month time horizon aggregated across six random seeds (exits counted after the warmup period). Subfigures (a) and (b) show bounded anchorage queues when arrival rates are below operating capacity, while subfigures (c) and (d) show unbounded queue growth beyond operating capacity. Subfigures (c) and (d) further motivate the concept of ultimate capacity, as vessel throughput into the port continues to increase despite unbounded queues.}
    \label{fig:scales}
\end{figure}

The distinction between operating and ultimate capacity can be illustrated by running the simulation under different vessel arrival rates and examining the resulting anchorage queue behavior. Figure~\ref{fig:scales} shows anchorage queue length over time for arrival rates of $0.7, 0.9, 1.1$, and $1.3$~vph over a six-month simulation horizon. The operating capacity of the port was estimated to be $0.9$~vph, as discussed in Section~\ref{sec:op_cap_results}. When the arrival rate is set equal to the operating capacity, spikes in the anchorage queue around 4{,}000 hours indicate a transition from stable to unstable system behavior. At higher arrival rates, the anchorage queue grows without bound, reflecting a highly unstable queueing behavior.

Despite this instability, the port processes a larger number of vessels due to the higher arrival volume. For example, 5{,}235 vessels are processed at an arrival rate of $\lambda = 1.3$~vph, compared to 3{,}832 vessels at the operating capacity over the same six-month horizon. This highlights the importance of estimating ultimate capacity: even under unstable conditions with unbounded queue growth, the system can sustain higher throughput over a finite time horizon. However, under such unstable conditions, the effects of the growing queue dominate, leading to ever-increasing delays for vessels arriving later.

To estimate the ultimate capacity of the Port of Houston, we ran the simulation under
different vessel arrival rates. To ensure that the relative proportions of vessel classes remain constant, arrival rates for each cargo class were scaled by a common factor ranging from $0.3$ to $4.6$ in increments of $0.1$, resulting in a total of 44 simulation runs. The simulation time horizon was held fixed across all runs. As arrival rates increased, we tracked the total number of vessel exits from the anchorage as a function of the arrival rate. The number of exits, excluding the warm-up period, was then fitted using the procedure described in Section~\ref{subsec:ul_cap}. The resulting ODE provided a good fit to the simulation data, with an root mean squared error (RMSE) of $33.63$ vessels over a six-month period. The estimated ultimate capacity of the port is $1.41$~vph. This estimate aligns closely with empirical studies of queue recovery at the Port of Houston following channel closures. For example, \cite{bathgateFog} report a mean recovery rate of approximately $0.8$~vph following fog-related channel closures at the Port of Houston over the period 2016--2024. This recovery rate reflects the net rate at which the anchorage queue dissipates once port operations resume. Given an average arrival rate of about $0.7$~vph during this period, the implied ultimate capacity can be inferred from the relationship that queue recovery occurs at a net rate of $(C_u - \lambda)$~vph. Equating this net recovery rate to the observed value of $0.8$~vph yields an implied ultimate capacity of $C_u \approx 0.8 + 0.7 = 1.5$~vph. This estimate is reasonably close to the ultimate capacity of $1.41$~vph obtained in this study. The remaining discrepancy is likely due to the substantial year-to-year variability in recovery rates reported by \cite{bathgateFog}.

\begin{table}
\centering
\begin{tabular}{l|c|c|c|c|c|c|c|c}
\hline
\textbf{Vessel type} 
& $\boldsymbol{\lambda}$ (vph)
& $\boldsymbol{C_o}$ (vph)
& $\boldsymbol{C_u}$ (vph)
& $\boldsymbol{C_s}$ (vph)
& $\boldsymbol{\theta}$ 
& $\boldsymbol{\beta}$ 
& $\boldsymbol{\lambda_c}$ (vph)
& \textbf{RMSE/$\mathbf{T}$} (vph) \\
\hline
All vessels
& 0.79 & 0.90 & 1.41 & 0.91 & 3.94 & 595.9 & 1.98 & 0.0078 \\
Container
& 0.10 & 0.15 & 0.31 & 0.080 & 7.45 & 39.1 & 0.25 & 0.0026\\
Non-container
& 0.14 & 0.18 & 0.56 & 0.53 & 8.16 & 0.62 & 0.57 & 0.0034\\
Tanker
& 0.55 & 0.57 & 0.86 & 0.33 & 6.03 & 425.7 & 1.11 & 0.0054\\
\hline
\end{tabular}
\caption{Estimated operating and ultimate capacity model parameters for the aggregate port system and individual cargo classes using simulation. ($\lambda$: arrival rate (vph);
$C_o$: operating capacity (vph);
$C_u$: ultimate capacity (vph);
$C_s$: saturated capacity (vph);
$\theta$: transition parameter for uncongested region;
$\beta$: transition parameter for congested region;
$\lambda_c$: vessel arrival rate at ultimate capacity (vph);
RMSE/T: root mean squared error of model fit divided by simulation time (vph)).} \label{tab:ul_params}
\end{table}

\begin{figure}
  \centering
  \begin{subfigure}[]{0.48\textwidth}
    \centering
    \includegraphics[width=\textwidth, trim=0 0 0 0, clip]{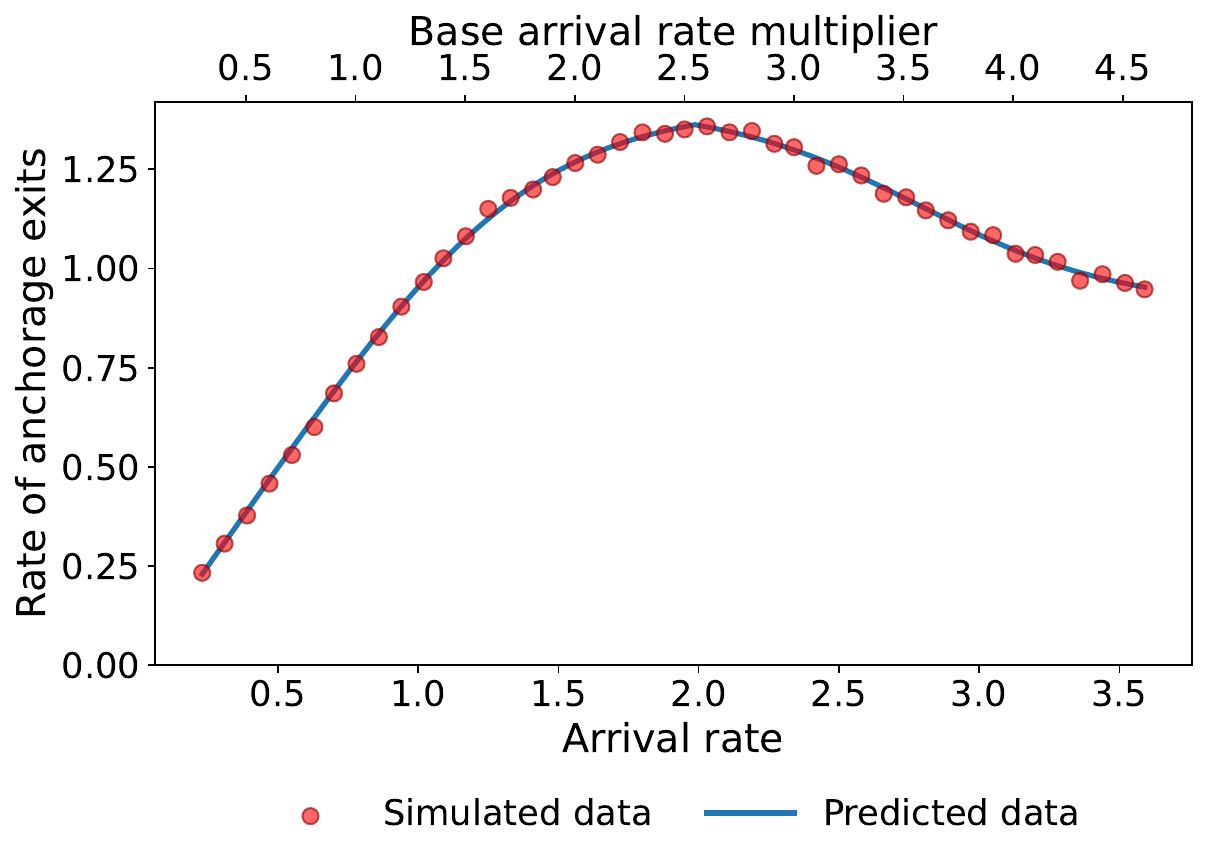}
    \caption{All vessels.}
    \label{fig:res_fit_a}
  \end{subfigure}
  \hfill
  \begin{subfigure}[]{0.48\textwidth}
    \centering
    \includegraphics[width=\textwidth, trim=0 0 0 0, clip]{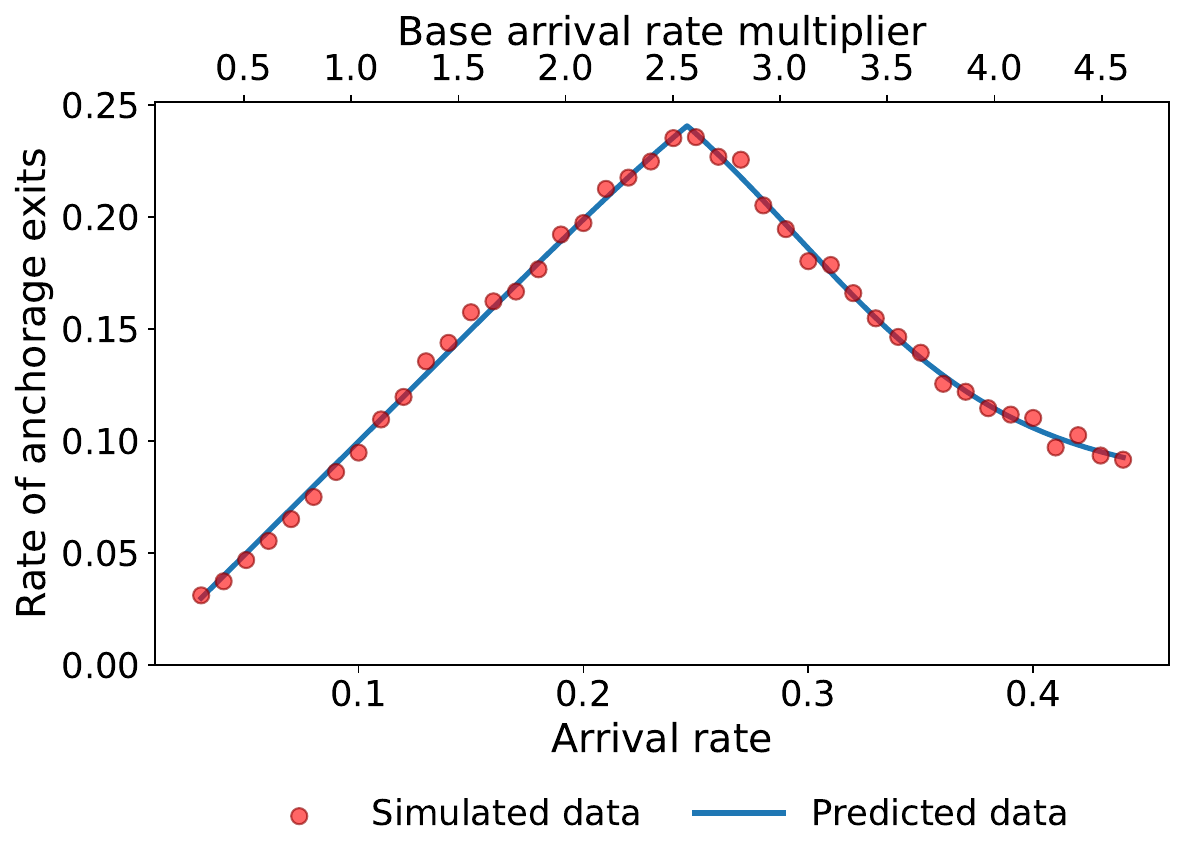}
    \caption{Container ships.}
    \label{fig:res_fit_b}
  \end{subfigure}
  
  \vspace{0.5cm}
  
  \begin{subfigure}[]{0.48\textwidth}
    \centering
    \includegraphics[width=\textwidth, trim=0 0 0 0, clip]{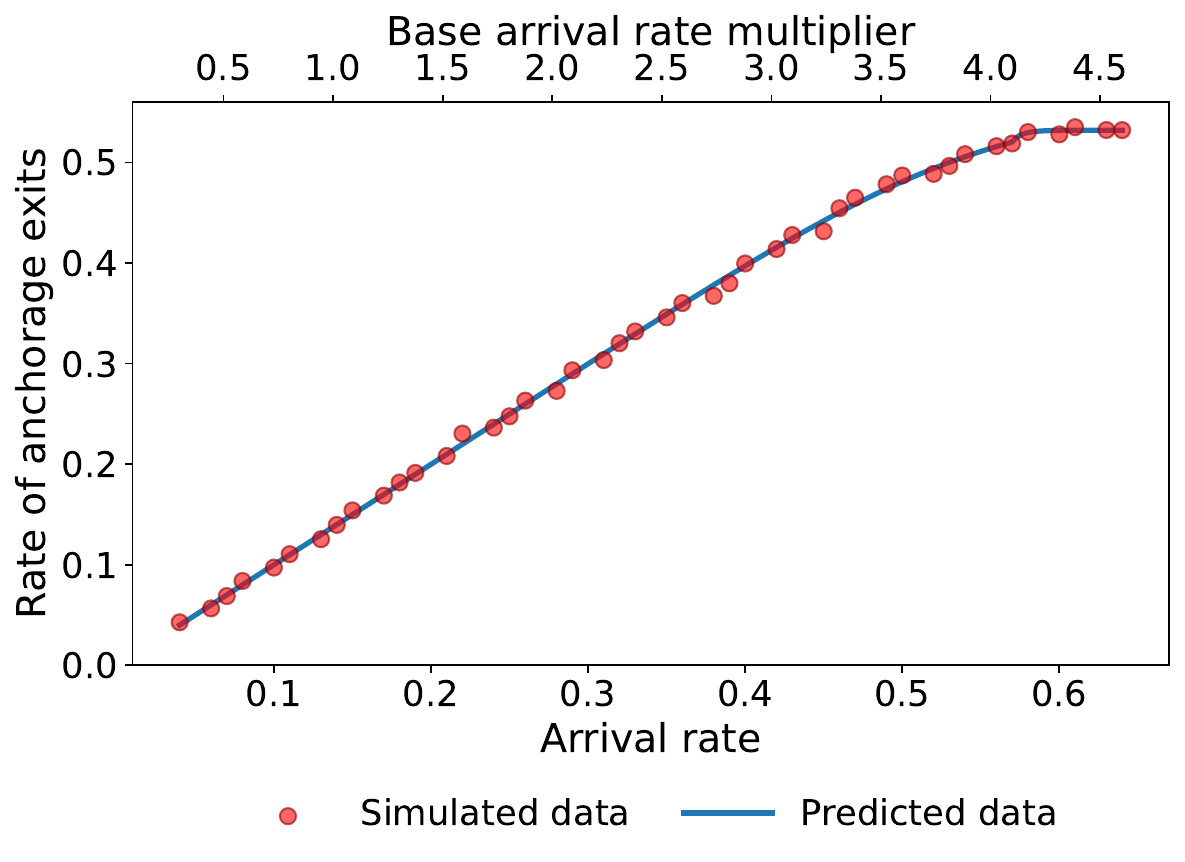}
    \caption{Non-container ships.}
    \label{fig:res_fit_c}
  \end{subfigure}
  \hfill
  \begin{subfigure}[]{0.48\textwidth}
    \centering
    \includegraphics[width=\textwidth, trim=0 0 0 0, clip]{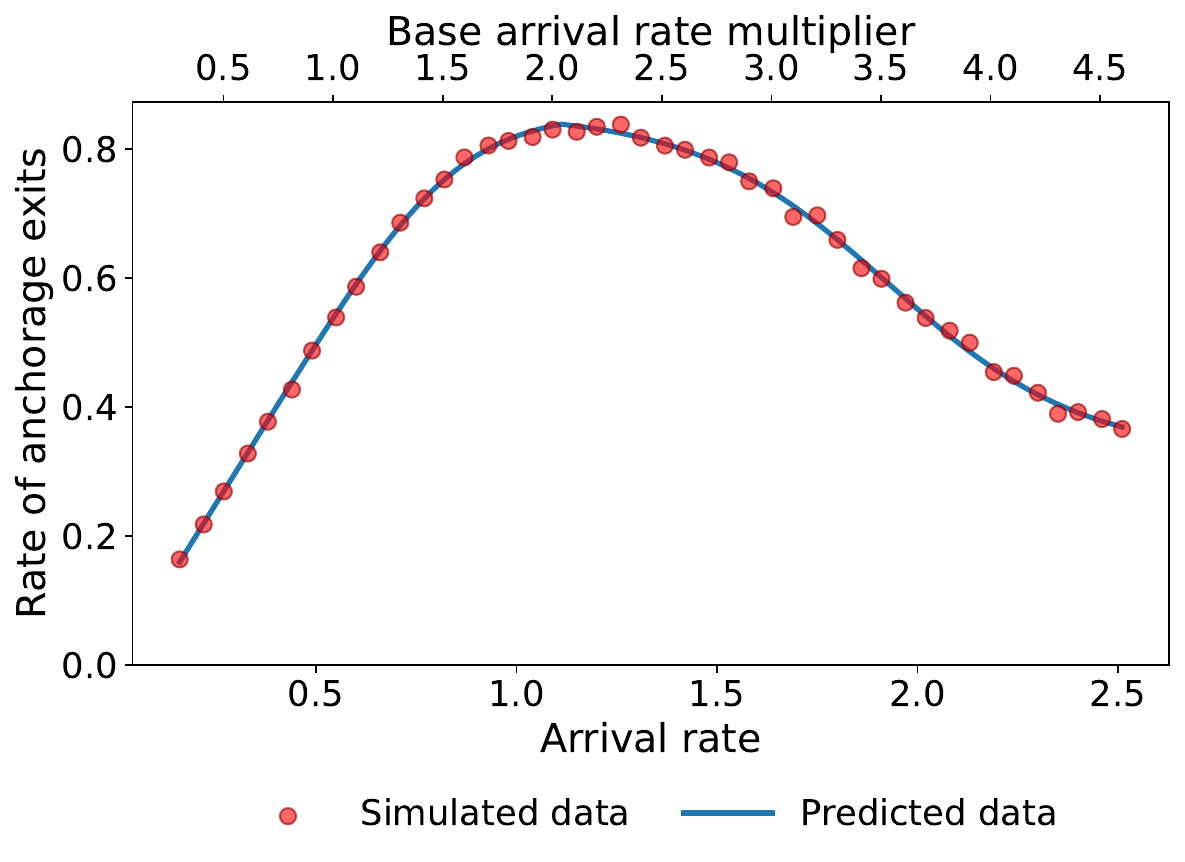}
    \caption{Tanker ships.}
    \label{fig:res_fit_d}
  \end{subfigure}
  
  \caption{ODE fit to simulation of different arrival rates for different vessel classes.}
  \label{fig:sctrtheo}
\end{figure}

We further conducted a class-wise analysis to estimate the ultimate capacities of individual cargo classes. For this analysis, the number of exits for each cargo class was recorded while
arrival rates for all vessels were increased proportionally. The ultimate capacities were estimated to be $0.31$, $0.56$, and $0.86$~vph for container, non-container, and tanker vessels, respectively. The corresponding RMSE values for these fits were $11.33$, $14.58$, and $23.44$ vessels respectively. Figure~\ref{fig:sctrtheo} shows the fitted curves together with the simulated data points. For these plots, the number of exits corresponds to a six-month period excluding the warm-up phase (i.e., $4320 - 1000 = 3320$ hours). The current arrival rate, the operating capacity obtained from simulation and the estimated model parameters ($C_u$, $C_s$, $\theta$, $\beta$, $\lambda_c$) for the aggregate system as well as for each cargo class are summarized in Table~\ref{tab:ul_params}.

\section{Bottleneck analysis and disruption planning}\label{sec:appnl} 

The models presented in this study can be used to estimate both the long-term operating capacity of the port system (Section~\ref{subsec:op_cap}) and the maximum achievable throughout, or ultimate capacity (Section~\ref{subsec:ul_cap}). These capacity measures have several practical applications, including bottleneck identification, infrastructure planning, and disruption management. This section illustrates how operating and ultimate capacity can be jointly applied to guide data-driven decisions regarding when and where to undertake capacity improvement projects.

Once a port simulation model is constructed, capacity analysis can be performed by systematically reducing individual resources and recording the resulting changes in operating and ultimate capacity. Resources to which these capacity measures are most sensitive are identified as critical bottlenecks in the port system. Importantly, reductions in different resources may affect operating capacity, ultimate capacity, or both. Resources whose reduction primarily lowers operating capacity correspond to limiting factors under normal operating conditions. Consequently, infrastructure investments aimed at improving long-term throughput should be prioritized toward these resources. In contrast, resources whose reduction predominantly lowers ultimate capacity are limiting under disrupted conditions. Strengthening these resources may improve post-disruption recovery, although such investments may not necessarily increase long-term throughput.

We illustrate this approach using a simulation framework for the Port of Houston. Multiple scenarios were examined by selectively reducing individual resources or their service rates by $10\%$ and $20\%$. Each scenario was simulated over a six-month horizon, with results averaged over six replicate runs. For the base case, the operating capacity was estimated at $0.90$~vph, while the ultimate capacity was $1.41$~vph. Given the current vessel arrival rate at the Port of Houston of $0.79$~vph, these estimates indicate that the port can sustain approximately a $14\%$ increase in arrival rates over extended periods without experiencing unbounded queue growth. The available slack is substantially higher over shorter time horizons, with the port able to accommodate up to $78.4\%$ more vessels over short periods. The operating and ultimate capacity from each scenario was then compared against the base capacities. 

\begin{table} \centering \begin{tabular}{p{4.5cm}cccc} \hline 
\textbf{Scenario} & $C_o$ (10\% change) & $C_o$ (20\% change) & $C_u$ (10\% change) & $C_u$ (20\% change)\\ \hline
\\
\textbf{Base Case} 
& \multicolumn{2}{c}{0.90} 
& \multicolumn{2}{c}{1.41} \\
\\
\textbf{Container terminals:} \\
Fewer berths & 0.90 (0.00\%) & 0.90 (0.00\%) & 1.41 (0.00\%) & 1.41 (0.00\%) \\
Lesser storage & 0.90 (0.00\%) & 0.90 (0.00\%) & 1.41 (0.00\%) & 1.41 (0.00\%) \\
Slower transfer & 0.90 (0.00\%) & 0.90 (0.00\%) & \textbf{1.40 (-0.71\%)} & \textbf{1.37 (-2.84\%)} \\
\\
\textbf{Non-containerized \newline cargo terminals:} \\
Fewer berths & 0.90 (0.00\%) & 0.90 (0.00\%) & 1.41 (0.00\%) & 1.41 (0.00\%) \\
Lesser storage & 0.90 (0.00\%) & 0.90 (0.00\%) & 1.41 (0.00\%) & 1.41 (0.00\%) \\
Slower transfer & 0.90 (0.00\%) & 0.90 (0.00\%) & 1.41 (0.00\%) & 1.41 (0.00\%) \\
\\
\textbf{Liquid-bulk terminals:} \\
Fewer berths & 0.90 (0.00\%) & \textbf{0.89 (-1.11\%)} & \textbf{1.36 (-3.55\%)} & \textbf{1.34 (-4.96\%)} \\
Lesser storage & 0.90 (0.00\%) & 0.90 (0.00\%) & 1.41 (0.00\%) & 1.41 (0.00\%) \\
Slower transfer & 0.90 (0.00\%) & \textbf{0.89 (-1.11\%)} & \textbf{1.37 (-2.84\%) }& \textbf{1.34 (-4.96\%)} \\
\\
\textbf{Pilots and tugboats:} \\
Fewer pilots & 0.90 (0.00\%) & 0.90 (0.00\%) & \textbf{1.31 (-7.09\%)} & \textbf{1.23 (-12.77\%)} \\
Fewer tugboats & 0.90 (0.00\%) & 0.90 (0.00\%) & 1.41 (0.00\%) & 1.41 (0.00\%) \\
\\
\textbf{Trucks and trains:} \\
Slower truck arrivals & 0.90 (0.00\%) & 0.90 (0.00\%) & 1.41 (0.00\%) & 1.41 (0.00\%) \\
Slower train arrivals & 0.90 (0.00\%) & 0.90 (0.00\%) & 1.41 (0.00\%) & 1.41 (0.00\%) \\
\hline \end{tabular} \caption{Bottleneck detection under normal and post-disruption scenarios over a six-month horizon.
($C_o$ (10\% change): operating capacity in vph under a 10\% resource perturbation;
$C_o$ (20\% change): operating capacity in vph under a 20\% resource perturbation;
$C_u$ (10\% change): ultimate capacity in vph under a 10\% resource perturbation;
$C_u$ (20\% change): ultimate capacity in vph under a 20\% resource perturbation;
percentages in parentheses indicate changes relative to the base case.)}
\label{tab:scenarios} \end{table}

The results each of the scenarios are summarized in Table~\ref{tab:scenarios}. Across all resources considered, at least a $10\%$ capacity slack is observed in terms of operating capacity. Under normal operating conditions, liquid-bulk terminal berths and pipelines emerge as the primary bottlenecks when these resources are constrained by $20\%$. Specifically, a $20\%$ reduction in either liquid-bulk termninal berth or pipeline transfer rates results in a $1.1\%$ decrease in operating capacity, whereas comparable reductions in other resources have negligible effects, indicating at least $20\%$ slack in those components. In contrast, ultimate capacity estimates reveal a different pattern. A $10\%$ reduction in pilot availability reduces ultimate capacity from $1.41$~vph to $1.31$~vph ($-7.1\%$), while a $20\%$ reduction lowers it further to $1.23$~vph ($-12.8\%$). Reductions in liquid-bulk terminal capacity also have a measurable impact on ultimate capacity: a $20\%$ decrease in either liquid-bulk terminal berths or transfer rates reduces ultimate capacity to $1.34$~vph ($-4.9\%$). Container terminal transfer rates and train arrivals also affect ultimate capacity, though to a lesser extent, with a $20\%$ reduction leading to decreases of $2.8\%$ and $0.7\%$, respectively. All other resources exhibit negligible changes in ultimate capacity under comparable perturbations, indicating substantial slack even under stressed conditions.

These results show that the binding capacity constraint changes across demand regimes. It is observed under regular operating conditions, capacity limitations are governed primarily by liquid-bulk terminal infrastructure. Accordingly, long-term capacity expansion efforts should prioritize investments in liquid-bulk terminal berths and transfer facilities. However, under high-demand or post-disruption scenarios, pilot availability emerges as the dominant bottleneck, particularly when arrival rates are elevated. While expanding pilot availability is unlikely to materially increase long-run throughput under typical conditions, it can substantially improve system resilience by accelerating recovery following disruptions.


\section{Conclusion} \label{sec:conc}

In this study, we define and present methods to compute operating and ultimate capacities for ports with an open anchorage and narrow waterway channels. Operating capacity refers to the maximum throughput attainable by the port over a long time horizon under stable anchorage queue formation, whereas ultimate capacity represents the short-term maximum physically attainable port throughput. We further show that while port throughput can temporarily exceed operating capacity, such operation leads to unstable anchorage queues and increased congestion for subsequently arriving vessels.

The two capacity measures have distinct implications for planning and decision-making. Operating capacity measurements are primarily used for long-term planning under stable conditions, while ultimate capacity measurements aid post-disruption planning and recovery analysis. We propose a queueing-theory-based model to estimate operating capacity for multi-modal port processes without relying on data-intensive simulation. On the other hand, our ultimate capacity metric relies on a simulation of the port. The simulation is run under varying arrival rates, and the resulting number of vessel exits is recorded. An ODE model is then fitted to the simulation-observed exit counts to estimate the ultimate capacity of the port. In addition to estimating ultimate capacity, the ODE formulation provides an exit-curve relationship that can be used to predict the maximum number of exits under varying arrival rates.

The proposed framework was validated using an extensive discrete-event simulation of the Port of Houston, calibrated with real-world archival AIS data, port record and practitioner interviews. The case study yielded the following conclusions:
\begin{itemize}
    \item The operating capacity of the Port of Houston is approximately $0.9$~vph, while the ultimate capacity is approximately $1.4$~vph.
    \item Under stable operating conditions, the channel restrictions are not binding constraints. The Houston Ship Channel capacity was found to be approximately $1.5$~vph.
    \item The terminal system limits throughput during normal operations. Liquid-bulk terminals are the most restrictive terminal group (highest utilization) within the terminal system.
    \item Following disruptions, pilot availability becomes the dominant system bottleneck. A $10\%$ decrease in pilot availability post-disruption decreases ultimate capacity by approximately $7.1\%$.
\end{itemize}

Nevertheless, the proposed capacity model has several limitations. First, simplifying assumptions were made regarding arrival processes, service time distributions, and queueing disciplines in our operating capacity model. Although Poisson arrivals and exponentially distributed service times are common assumptions in queueing theory, empirical evidence suggests that seaport operations may deviate from these assumptions  \citep{legato2020queueing}. In addition, the capacity model assumed a FCFS queueing discipline for all vessels, which may not hold in practice. Extending the proposed framework to accommodate more general arrival processes, service distributions, and priority rules is a promising direction for future research.

Furthermore, while the presented models can estimate ultimate capacity accurately, it does not quantify how long such capacity levels can be sustained. Future work is needed to characterize this duration, either analytically or empirically. Additionally, the accuracy of ultimate capacity measurements depends on how well the underlying simulation models the port under consideration. An promising direction is to examine how input parameter uncertainty could be dealt in a more robust fashion. 

Finally, the estimated ultimate-capacity exit curves provide a direct relationship between arrival rates and vessel exits, making the framework particularly well suited for detailed post-disruption analyses with time-varying arrivals. In the Port of Houston case, such analyses can be conducted without additional simulation by relying solely on the exit curves presented in this study, providing a promising direction for future research.

\section*{Acknowledgments}
This work was partially funded through U.S. Army Engineer Research and Development Center contract \#W912HZ23C0052, ``Data-Driven, Multimodal Freight Modeling for Waterways and Ports." The findings of this study reflect the authors' work and are not official policies or standards of the U.S. Army Corps of Engineers. This work was also partially supported by the University Transportation Center National Center for Understanding Future Travel Behavior and Demand. The authors thank Mark A. Cowan of ERDC for his valuable input and feedback that guided this work.

\section*{Declaration of generative AI and AI-assisted technologies in the writing process}
During the preparation of this work, the authors used GPT-5.2 in order to improve writing in select parts of the paper. After using this tool/service, the authors reviewed and edited the content as needed and take full responsibility for the content of the published article.

\section*{Author Contributions}

The authors confirm their contribution to the paper as follows:\\
\textbf{Conceptualization:} Debojjal Bagchi, Kyle Bathgate\\
\textbf{Methodology:} Debojjal Bagchi, Kyle Bathgate\\
\textbf{Formal analysis:} Debojjal Bagchi, Kyle Bathgate, Stephen D. Boyles\\
\textbf{Data curation:} Debojjal Bagchi, Kyle Bathgate, Magdalena I. Asborno, Marin M. Kress\\
\textbf{Validation:} Debojjal Bagchi, Kyle Bathgate, Kenneth N. Mitchell, Magdalena I. Asborno, Marin M. Kress\\
\textbf{Writing (original draft):} Debojjal Bagchi\\
\textbf{Writing (review \& editing):} Debojjal Bagchi, Stephen D. Boyles, Kyle Bathgate, Kenneth N. Mitchell, Magdalena I. Asborno, Marin M. Kress\\
\textbf{Supervision:} Stephen D. Boyles\\
\textbf{Project administration:} Kenneth N. Mitchell, Magdalena I. Asborno, Marin M. Kress\\
\textbf{Funding acquisition:} Kyle Bathgate, Stephen D. Boyles\\
All authors reviewed the results and approved the final version of the manuscript.

\bibliographystyle{elsarticle-num-names} 
\bibliography{references}

\end{document}

%% file: header.tex
\usepackage{algorithm}
\usepackage[noend]{algpseudocode}  
\usepackage{makecell} 
\usepackage{amsmath}
\usepackage{subcaption} 
\usepackage{amsthm}
\usepackage{amssymb}
\usepackage{array}
\usepackage{bm}
\usepackage{enumerate}
\usepackage{fancyhdr}
\usepackage[pdftex]{graphicx}
\usepackage{multirow}
\usepackage{nag}
\usepackage{natbib}
\usepackage{verbatim}
\usepackage[colorlinks]{hyperref}
\usepackage{url}
\usepackage[dvipsnames]{xcolor}
\usepackage{tcolorbox}
\usepackage{tabularx}
\usepackage{pgfplots}
\usepackage{etoolbox}

\hypersetup{bookmarks=true, citecolor=blue, urlcolor=blue, linkcolor=blue}
\usepackage[left=0.75in,right=0.75in,top=0.75in,bottom=0.75in]{geometry}

\usepackage{titlecaps}
\usepackage[utf8]{inputenc}
\usepackage{amsmath}
\usepackage{amsfonts}
\usepackage{amssymb}
\usepackage{algorithm}
\usepackage[noend]{algpseudocode}
\usepackage{multirow}
\usepackage{float}
\usepackage{array}
\usepackage{lipsum}
\usepackage{comment}
\usepackage{graphicx}
\usepackage{url}
\usepackage{natbib}
\usepackage{color}
\usepackage{pgfplots}
\usepackage{mathtools}
\usepackage{caption}
\usepackage{subcaption}
\usepackage{pdflscape}
\usepackage{booktabs} 
\usepackage{enumerate}
\usepackage{authblk}
\usepackage{blkarray}
\usepackage{bm}
\usepackage{marvosym} 
\usepackage{soul}
\usepackage{textcomp, xspace}
\usepackage{lineno}

\usepackage{longtable}
\newtheorem{dfn}{Definition}

\definecolor{greenxllite}{RGB}{196,215,155}
\definecolor{bluexllite}{RGB}{149,179,215}
\definecolor{redxllite}{RGB}{218,150,148}
\definecolor{greenxl}{RGB}{155, 187, 89}
\definecolor{bluexl}{RGB}{79,129,189}
\definecolor{redxl}{RGB}{192, 80, 77}
\definecolor{OliveGreen}{RGB}{70,136,52}